\documentclass[useAMS,usenatbib,usegraphicx]{mn2e}
\usepackage{bm}
\usepackage{psfig}
\usepackage{amssymb}
\usepackage{amsmath}
\usepackage{multirow}
\usepackage{twoopt}
\usepackage{times}
\usepackage{color}
\usepackage{units}
\usepackage{subfigure}
\usepackage{version}

\usepackage[pdftex,             
            breaklinks=true,%
            colorlinks=true,%
            citecolor=black,
            pdfauthor={Bonnivard et al.},%
            pdftitle={Template for manuscripts MNRAS}%
           ]{hyperref}

\makeatletter

\newcommand{\beq}{\begin{equation}}
\newcommand{\eeq}{\end{equation}}
\newcommandtwoopt{\citeads}[3][][]{\href{http://adsabs.harvard.edu/abs/#3}{\def\hyper@linkstart##1##2{}\let\hyper@linkend\@empty\citealpads[#1][#2]{#3}}}
\newcommandtwoopt{\citepads}[3][][]{\href{http://adsabs.harvard.edu/abs/#3}{\def\hyper@linkstart##1##2{}\let\hyper@linkend\@empty\citep[#1][#2]{#3}}}
\newcommandtwoopt{\citetads}[3][][]{\href{http://adsabs.harvard.edu/abs/#3}{\def\hyper@linkstart##1##2{}\let\hyper@linkend\@empty\citet[#1][#2]{#3}}}
\newcommandtwoopt{\citealpads}[3][][]{\href{http://adsabs.harvard.edu/abs/#3}{\def\hyper@linkstart##1##2{}\let\hyper@linkend\@empty\citealp[#1][#2]{#3}}}
\newcommandtwoopt{\citealtads}[3][][]{\href{http://adsabs.harvard.edu/abs/#3}{\def\hyper@linkstart##1##2{}\let\hyper@linkend\@empty\citealt[#1][#2]{#3}}}
\newcommandtwoopt{\citetadsads}[3][][]{\href{http://adsabs.harvard.edu/abs/#3}{\def\hyper@linkstart##1##2{}\let\hyper@linkend\@empty\citetads[#1][#2]{#3}}}
\newcommandtwoopt{\citealpadsads}[3][][]{\href{http://adsabs.harvard.edu/abs/#3}{\def\hyper@linkstart##1##2{}\let\hyper@linkend\@empty\citealpads[#1][#2]{#3}}}
\newcommandtwoopt{\citealtadsads}[3][][]{\href{http://adsabs.harvard.edu/abs/#3}{\def\hyper@linkstart##1##2{}\let\hyper@linkend\@empty\citealtads[#1][#2]{#3}}}
\newcommandtwoopt{\citeyearads}[3][][]{\href{http://adsabs.harvard.edu/abs/#3}{\def\hyper@linkstart##1##2{}\let\hyper@linkend\@empty\citeyear[#1][#2]{#3}}}
\newcommandtwoopt{\citeadsstar}[3][][]{\href{http://adsabs.harvard.edu/abs/#3}{\def\hyper@linkstart##1##2{}\let\hyper@linkend\@empty\citealpads*[#1][#2]{#3}}}
\newcommandtwoopt{\citepadsstar}[3][][]{\href{http://adsabs.harvard.edu/abs/#3}{\def\hyper@linkstart##1##2{}\let\hyper@linkend\@empty\citep*[#1][#2]{#3}}}
\newcommandtwoopt{\citetadsadsstar}[3][][]{\href{http://adsabs.harvard.edu/abs/#3}{\def\hyper@linkstart##1##2{}\let\hyper@linkend\@empty\citetads*[#1][#2]{#3}}}
\newcommandtwoopt{\citeyearadsstar}[3][][]{\href{http://adsabs.harvard.edu/abs/#3}{\def\hyper@linkstart##1##2{}\let\hyper@linkend\@empty\citeyear*[#1][#2]{#3}}}
\newcommandtwoopt{\citeauthoradsstar}[3][][]{\href{http://adsabs.harvard.edu/abs/#3}{\def\hyper@linkstart##1##2{}\let\hyper@linkend\@empty\citeauthor*[#1][#2]{#3}}}
\newcommandtwoopt{\citepthesis}[3][][]{\href{http://tel.archives-ouvertes.fr/docs/#3}{\def\hyper@linkstart##1##2{}\let\hyper@linkend\@empty\citep[#1][#2]{#3}}}
\newcommandtwoopt{\citetadsthesis}[3][][]{\href{http://tel.archives-ouvertes.fr/docs/#3}{\def\hyper@linkstart##1##2{}\let\hyper@linkend\@empty\citetads[#1][#2]{#3}}}
\makeatother
\def\aap{A\&A}%
\def\apjl{ApJ}%
\def\apjs{ApJS}%
\def\ao{Appl.~Opt.}%
\begin{document}

\title[Contamination in dSphs: Segue~I]{Contamination of stellar-kinematic samples and uncertainty about dark matter annihilation profiles in ultrafaint dwarf galaxies: the example of Segue~I}

\author[Bonnivard, Maurin, Walker]{V. Bonnivard$^{1}$\thanks{E-mails:bonnivard@lpsc.in2p3.fr (VB), dmaurin@lpsc.in2p3.fr (DM), mgwalker@andrew.cmu.edu (MGW)}, D. Maurin$^{1}$, M. G. Walker$^{2,3}$ \\
  $^1$LPSC, Universit\'e Grenoble-Alpes, CNRS/IN2P3,
     53 avenue des Martyrs, 38026 Grenoble, France\\
  $^2$Department of Physics, Carnegie Mellon University, Pittsburgh, PA 15213, USA\\
  $^3$McWilliams Center for Cosmology, 5000 Forbes Avenue Pittsburgh, PA 15213, USA\\
}

\pagerange{\pageref{firstpage}--\pageref{lastpage}} \pubyear{Xxxx}
\date{Accepted Xxxx. Received Xxxx; in original form Xxxx}
\label{firstpage}

\maketitle
\begin{abstract}
The expected gamma-ray flux coming from dark matter annihilation in dwarf spheroidal (dSph) galaxies depends on the so-called `$J$-factor', the integral of the squared dark matter density along the line-of-sight.  We examine the degree to which estimates of $J$ are sensitive to contamination (by foreground Milky Way stars and stellar streams) of the stellar-kinematic samples that are used to infer dark matter densities in `ultrafaint' dSphs.  Applying standard kinematic analyses to hundreds of mock data sets that include varying levels of contamination, we find that mis-classified contaminants can cause $J$-factors to be overestimated by orders of magnitude.  Stellar-kinematic data sets for which we obtain such biased estimates tend 1) to include relatively large fractions of stars with ambiguous membership status, and 2) to give estimates for $J$ that are sensitive to specific choices about how to weight and/or to exclude stars with ambiguous status. Comparing publicly-available stellar-kinematic samples for the nearby dSphs Reticulum~II and Segue~I, we find that only the latter displays both of these characteristics.  Estimates of Segue~I's $J$-factor should therefore be regarded with a larger degree of caution when planning and interpreting gamma-ray observations. Moreover, robust interpretations regarding dark matter annihilation in dSph galaxies in general will require explicit examination of how interlopers might affect the inferred dark matter density profile.
\end{abstract}


\section{Introduction}

Owing to their proximity, large dynamical mass-to-light ratios and low astrophysical backgrounds \citepads{1983ApJ...266L..11A,1998ARA&A..36..435M,2012AJ....144....4M}, the Milky Way's dwarf spheroidal (dSph) satellites have become popular targets in `indirect' searches for particle dark matter (DM) via observations of high-energy photons that may be produced in annihilation events (e.g., \citealpads{1990Natur.346...39L,2004PhRvD..69l3501E,2013PhR...531....1S}).  In order to translate a given photon flux (or non-detection) into constraints (or upper limits) on cross-sections for DM annihilation, one must estimate the density of DM in the source.  This necessity helps to motivate observations and dynamical analyses of the tiny stellar populations that, for most of these galaxies, represent the only viable tracers of gravitational potentials. 

Stellar populations within the Milky Way's dSph companions are supported against gravity by random motions---i.e., they do not rotate.  Therefore, dynamical inferences about their gravitational potentials must appeal to statistical analyses of distributions of stellar positions and line-of-sight (l.o.s.) velocities.  For the most luminous, `classical' dSphs, estimates of DM densities are based on kinematic samples containing hundreds to thousands of stars per galaxy \citepads{2009AJ....137.3100W}, giving relatively tight constraints \citepads{2008Natur.454.1096S,2011MNRAS.418.1526C,2012PhRvD..86b3528C,2015ApJ...801...74G,2015MNRAS.453..849B}.  However, many efforts at indirect detection focus on the least luminous, `ultrafaint' satellites, for which stellar-kinematic sample sizes are necessarily smaller, ranging from single to double figures \citepads{2007MNRAS.380..281M,2007ApJ...670..313S}.  Despite larger uncertainties about their DM halos, the extreme properties of `ultrafaints' can give them extraordinary power in constraining properties of the DM particle \citepads{2011PhRvL.107x1303G,2011PhRvL.107x1302A,2015PhRvD..91h3535G,2014PhRvD..89d2001A}.  

Among the `ultrafaint' dSphs, Segue~I (hereafter `Seg~I', \citealtads{2007ApJ...654..897B}) and the newly-discovered Reticulum II (`Ret~II', \citealtads{2015ApJ...805..130K,2015ApJ...807...50B}) have become objects of particular interest in the search for gamma-rays from DM annihilation.  These objects combine the virtues of 1) large inferred DM densities (dynamical masses enclosed within their central $\sim 30$ pc are $\sim 10^5M_{\odot}$) and 2) relatively small distances ($d\sim 30$ kpc).  As such, these objects offer perhaps the best opportunities either to detect photons from DM annihilation\footnote{Based on public data from the Fermi Large Area Telescope, \citetads{2015PhRvL.115h1101G} report an excess of emission in the direction of Ret~II at energies $\sim 2-10$ GeV with significance $\ga 2.3\sigma$.  Based on proprietary data, the \citetads{2015ApJ...809L...4D} finds an excess at the same energies in Ret II, but assigns the signal significance of just $0.3\sigma$.  Revisiting the public Fermi-LAT data, \citetads{2015JCAP...09..016H} confirm the detection of \citetads{2015PhRvL.115h1101G} and assign it similar significance.} or, in the absence of a signal, place stringent upper limits on cross-sections for various annihilation channels.  

Given their propagation into constraints on particle physics, it is important to examine robustness of dynamical inferences about dSph DM densities.  In previous work, we have used mock stellar-kinematic data sets to examine the sensitivity of dynamical analyses to modelling assumptions such as spherical symmetry, parametrisations of DM and stellar density profiles, and behaviour of the velocity anisotropy profile \citepads{2015MNRAS.446.3002B}.  In general, we find that violations of simplistic modelling assumptions give rise to systematic errors that dominate random errors only for the larger samples characteristic of `classical' dSphs.  For `ultrafaints', such systematics are well contained within the relatively large statistical errors associated with smaller kinematic samples.  

The conclusions from our previous work hold at least when samples are free from contamination by non-member stars, as were the mock data sets analysed therein.  However, given what is often substantial overlap among distributions of observables (e.g., position, velocity, chemical abundances), even the largest and most precise stellar-kinematic data sets do not let us identify with absolute certainty which stars belong to the dwarf galaxy and which are contaminants in the Galactic foreground.  

Here, we extend our previous work to consider the impact of sample contamination on estimates of dSph DM densities and the ensuing expectations for indirect detection of DM particles.  Specifically, we test the reliability of estimates of the DM annihilation factor $J$ --- the l.o.s. integral of the squared DM density (Section~\ref{sec:jeans}) ---
on a new suite of mock data sets that include varying degrees of contamination by populations that mimic Galactic foreground as well as tidal streams in which the dSph may be embedded (Section~\ref{sec:mock}).  Applying analyses that are similar to those used to infer DM densities for real dSphs, we examine the severity of systematic errors introduced by sample contamination, and we identify characteristic features of data sets that are particularly susceptible to such errors (Section~\ref{sec:mock_tests}).  Finally, we use these results to gauge the reliability of dynamically-inferred $J$-factors for the specific dSphs Seg~I and Ret~II (Section~\ref{sec:seg1}). We stress that this work relies on the use of a large number of mock data sets, requiring the analysis to be fast. For this reason, the results presented in this study rely on a standard Jeans analysis in which the contamination estimation and the DM profiles reconstruction are dealt with in two separate steps. An alternative Bayesian analysis, using a global likelihood which allows to constrain simultaneously the two, is presented in Appendix~\ref{subsubsec:mixture}. However, the relative computational expense of this method inhibits the extensive testing to which we have subjected the more standard analyses employed in this and previous work.  In the few cases where we have tested it and can compare directly against the performance of the standard analyses, we find no significant improvement (e.g., in heavily-contaminated mock data sets where the standard analyses give strongly biased results for $J$-factors, we find similarly strong biases using the global likelihood). As summarized in Section~\ref{sec:conc}, this indicates that a two step analysis is sufficient for assessing contamination impact on $J$-factors. Improving the algorithms used to separate members from contaminant in dSphs, and/or getting more data should be the priority to obtain more robust constraints on contaminated objects such as Segue~I.

\section{Estimation of dark matter density and the $J$-factor}
\label{sec:jeans}

The differential $\gamma$-ray flux received on Earth in solid angle $\Delta\Omega$ from DM annihilation is related to the DM particle mass, the thermally-averaged annihilation cross section $\langle \sigma v\rangle$, the energy spectrum of annihilation products, and the quantity
\beq
J(\Delta\Omega)=\int_{\Delta\Omega}\int \rho^2_{\rm DM} (l,\Omega) \,dld\Omega\;.
\label{eq:J}
\eeq 
The `$J$-factor', calculated here with the public code {\sc clumpy}\footnote{\tt \url{http://lpsc.in2p3.fr/clumpy}} \citepads{2012CoPhC.183..656C,2016CoPhC.200..336B}, is the integral of the squared DM density along l.o.s. $l$ and over solid angle $\Delta\Omega = 2 \pi\times[1-\cos{\alpha_{\rm int}}]$, with $\alpha_{\rm int}$ the angle over which the $\gamma$-ray signal is observed.  For a given particle physics model and measurement of $\gamma$-ray flux, the ensuing constraints on DM properties depend directly on $J$. For dSph galaxies, the $J$-factor is best constrained at the critical integration angle $\alpha_c = 2\times r_h/d$, where $r_h$ is the half-light radius and $d$ is the distance to the dSph \citepads{2011ApJ...733L..46W,2011MNRAS.418.1526C}.

\subsection{Jeans analysis}
Following many previous analyses (e.g., \citealtads{2007PhRvD..75h3526S,2010PhRvD..82l3503E,2011MNRAS.418.1526C}), here we estimate DM density profiles, and hence $J$-factors, by modelling stellar-kinematic data under assumptions of spherical symmetry, dynamical equilibrium and negligible rotational support.  The Jeans equation then relates observables (stellar positions and velocities) to the gravitational potential: \citepads{2008gady.book.....B}
\begin{equation}
\frac{1}{\nu}\frac{d}{dr}(\nu \bar{v_r^2})+2\frac{\beta_{\rm ani}(r)\bar{v_r^2}}{r}=-\frac{GM(r)}{r^2}.
\label{eq:jeans}
\end{equation}
Here, $\nu(r)$ is the stellar number density profile, $\bar{v_r^2}(r)$ is the radial velocity dispersion profile, $\beta_{\rm ani}(r)\equiv 1-\bar{v_{\theta}^2}/\bar{v_r^2}$ is the anisotropy profile of the stellar velocity dispersion, and $M(r)$ the mass---assumed to be dominated by DM---enclosed within radius $r$.  Current observations resolve neither internal proper motions nor relative distances of stars in a given dSph, confining empirical information to three dimensions of phase space (projected position and l.o.s. velocity). After solving Eq.~(\ref{eq:jeans}) and projecting along the l.o.s., the (squared) stellar velocity dispersion at projected radius $R$ reads
\begin{equation}
  \sigma_p^2(R)=\frac{2}{I(R)}\displaystyle \int_{R}^{\infty}\biggl (1-\beta_{\rm ani}(r)\frac{R^2}{r^2}\biggr ) 
  \frac{\nu(r)\, \bar{v_r^2}(r)\,r}{\sqrt{r^2-R^2}}\mathrm{d}r,
  \label{eq:jeansproject}
\end{equation}
with $I(R)$ the projected stellar number density profile.

Given a stellar-kinematic data set, we compare the l.o.s. stellar velocities to a model velocity dispersion, $\sigma_p(R)$, computed using parametric forms for the velocity anisotropy $\beta_{\rm ani}(r)$ and DM density profile $\rho_{\rm DM}(r)$ \citepads{2015MNRAS.446.3002B}.  For the DM density profile, $\rho_{\rm DM}(r)$, we adopt the Einasto profile:
\begin{equation}
  \rho_{\rm DM}(r)=\rho_{-2}\exp\biggl\{-\frac{2}{\alpha}\biggl [\biggl (\frac{r}{r_{-2}}\biggr)^{\alpha}-1\biggr ]\biggr \},
  \label{eq:einasto}
\end{equation}
where $\alpha$ controls the logarithmic slope, $r_{-2}$ is a scale radius and $\rho_{-2}$ is a scale density.  

We consider real data for known dSphs as well as mock data for which the true DM density profile is known. 
Before fitting the kinematic data for real dSphs, we estimate $I(R)$ by fitting a model of the form
\begin{equation}
  I_{\rm model}(R)=2\displaystyle\int_{R}^{\infty}\frac{\nu(r)rdr}{\sqrt{r^2-R^2}}+I_{\rm bkd},
\end{equation}
---i.e., the sum of the l.o.s. projection of $\nu(r)$ and a uniform background density---to the unbinned distribution of positions of stars identified as red giant candidates in photometric survey data.  Following \citetads[][see their Section 2 for details]{2015MNRAS.446.3002B}, we model the stellar density profile as a broken power law:
\begin{equation}
  \nu(r)=\frac{\nu_s^{\star}}{(r/r_s^{\star})^{\gamma^{\star}}[1+(r/r_s^{\star})^{\alpha^{\star}}]^{(\beta^{\star}-\gamma^{\star})/\alpha^{\star}}},
  \label{eq:nu}
\end{equation}
where $\alpha^{\star}$ controls the sharpness of transition from inner log-slope $\gamma^{\star}$ to outer log-slope $\beta^{\star}$ at scale radius $r_s^{\star}$.  When analysing mock data sets, we adopt the true profiles for $\nu(r)$, $I(R)$, and $\beta_{\rm ani}(r)$, in order to isolate systematic errors arising from contamination of stellar-kinematic samples.  For the two real `ultrafaint' dSphs considered in Section \ref{sec:seg1}, we assume  constant anisotropy, $\beta_{\rm ani}=\beta_0$, but have confirmed that the results would be similar if we used more flexible models for $\beta_{\rm ani}$ (e.g. Baes \& van Hese profile \citepads{2007A&A...471..419B}, see \citealtads{2015MNRAS.453..849B}).    

We estimate the free parameters using either a Markov Chain Monte Carlo (MCMC) engine, or a $\chi^2$ minimisation algorithm with bootstrap resampling (see Section \ref{subsec:bootstrap}). Thus, our fits have three free parameters ($\rho_{-2}$, $r_{-2}$ and $\alpha$) when applied to mock data, and four (now including a constant anisotropy $\beta_{0}$) when applied to real data.  For all parameters we adopt the same priors as \citetads[][see their Table 1]{2015MNRAS.453..849B}.  

\subsection{MCMC analysis: $J_{\rm median}$ and credible intervals}
\label{subsec:MCMC}
Following \citetads{2015MNRAS.453..849B,2015ApJ...808L..36B}, we use the {\tt GreAT} MCMC engine \citepads{2014PDU.....5...29P} to efficiently sample the multi-dimensional parameter space.  This procedure returns posterior probability density functions (PDFs) for model parameters.  Via Eq. (\ref{eq:J}), we can compute median values and credible intervals (CIs) for $J$-factors at any integration angle (e.g., \citealtads{2011MNRAS.418.1526C}).

For our MCMC analysis, we adopt a likelihood function that includes stellar velocity measurements as discrete data points ~\citepads{2008ApJ...678..614S,2009JCAP...06..014M}\footnote{An alternative method would be to bin the velocites in order to estimate the observed velocity dispersion profile, $\sigma_{\rm obs}(R)$, and compare it to the model $\sigma_p(R)$ \citepads{2009ApJ...704.1274W,2011MNRAS.418.1526C}. Using large sets of mock data, \citetads{2015MNRAS.453..849B} found that the two methods give similar results.}.  Rather than discarding observations from stars suspected of being contaminants, we can include all data and weight the $i^{\rm th}$ observation by a probability of membership, $P_i$, that can be estimated from observables under specific assumptions about member and contaminant populations (e.g., \citealtads{2007MNRAS.378..353K,2007MNRAS.377..843W,2009AJ....137.3109W,2009JCAP...06..014M}).  We adopt a likelihood function that has natural logarithm
\begin{eqnarray}
\log \mathcal{L}\!=-\frac{1}{2}\displaystyle\sum_{i=1}^{N_{\rm stars}}P_i\log (2\pi)-\frac{1}{2}\displaystyle\sum_{i=1}^{N_{\rm stars}}P_i\log (\sigma^2_p(R_i)+\Delta^2_{v_i})\nonumber\\
  -\frac{1}{2}\displaystyle\sum_{i=1}^{N_{\rm stars}}\frac{P_i(v_i-\langle v\rangle)^2}{\sigma^2_p(R_i)+\Delta^2_{v_i}},
  \label{eq:likelihood}
\end{eqnarray}
i.e. we assume that the distribution of l.o.s. stellar velocities $v$ is Gaussian, centred on the mean stellar velocity $\langle v \rangle$, and that the dispersion of velocities at radius $R$ comes from both the intrinsic dispersion $\sigma_p(R)$ and the measurement uncertainty $\Delta_{v}$.

\subsection{Bootstrap analysis: $\langle J\rangle_{\rm bootstrap}$ and $\sigma_{\rm bootstrap}$}
\label{subsec:bootstrap}
We also use in this work a bootstrap estimator \citepads{1982jbor.book.....E} for the $J$-factor and its statistical uncertainty. For a given kinematic data sample, we generate $N=500$ resamples by drawing, with replacement, $n$ stars from the $n$ observations of the original data set. For each resample, we find the best-fitting model using a $\chi^2$ minimisation algorithm, with the $\chi^2$ function
\begin{equation}
\chi^2 = -2 \log{\mathcal{L}},
\end{equation}
$\log\mathcal{L}$ being the log-likelihood from Eq.~(\ref{eq:likelihood}). We then compute the $J$-factor of each resample. The mean and dispersion over the $N=500$ values are the bootstrap estimators of the $J$-factor ($\langle J\rangle_{\rm bootstrap}$) and its statistical uncertainty ($\sigma_{\rm bootstrap}$).

The motivation for this approach is that it can reveal sensitivity to the presence of a few outliers in cases where sample contamination is important. The estimated $J$-factor might change drastically from one realization to the next, depending on whether or not the contaminants are drawn, and this can result in larger statistical uncertainties (than the associated MCMC-derived uncertainties) on the $J$-factor.

\section{Mock data description}
\label{sec:mock}
In order to investigate the impact on estimates of $J$-factors due to contamination of stellar-kinematic samples by non-member stars, we generate thousands of mock data sets that each sample mixtures of three simulated stellar populations tracing a gravitational potential dominated by DM.  

\subsection{Simulated populations}
\label{subsec:simpops}

The first simulated stellar population represents bona fide members of an `ultrafaint' (resp. `classical')-like dwarf galaxy\footnote{The `classical'-like mock dSphs are studied in \citetads{2015MNRAS.453..849B} (see their Appendix C), and will not be further discussed in this paper.}.  For this population, following the procedure described by \citetads{2011MNRAS.418.1526C}, we draw phase-space coordinates randomly from distribution functions that satisfy the collisionless Boltzmann equation (\citealtads{2008gady.book.....B}, see Section \ref{subsec:dfs} for details). The second population represents contamination from a tidal stream in which the `ultrafaint' may be embedded, as may be the case if the object formed as the satellite of a more massive and less dense `classical' dwarf that was more easily disrupted by Galactic tides \citepads{2009MNRAS.397.1748B}. The third population represents contaminant stars in the Galactic foreground, which often have the same colours and magnitudes as the red giants targeted in spectroscopic surveys of dSph galaxies. 

In order to generate a given mock stellar-kinematic data set, we first determine the contribution from each stellar population by drawing three random numbers.  The first, drawn from the interval $\log_{10} N_{\rm min}\leq \log_{10} N\leq \log_{10} N_{\rm max}$ with uniform probability, sets the overall sample size, $N$. For realisations corresponding to `ultrafaint' (resp. `classical') dSphs, we adopt $\log N_{\rm min}=1.5$ and $\log N_{\rm max}=2.0$ (resp. $\log N_{\rm min}=2.5$, $\log N_{\rm max}=3.5$). The second random number, drawn from the interval $0.1\leq f_{1}\leq 0.9$ with uniform probability, gives the fraction of $N$ stars drawn from either the member or stream population. The third, drawn from the interval $0.1\leq f_{2}\leq 0.9$ with uniform probability, gives the fraction of those $f_1N$ stars that are drawn from the member population. Thus the numbers of stars drawn from member, stream and Galactic foreground populations are $Nf_1f_2$, $Nf_1(1-f_2)$ and $N(1-f_1)$, respectively.

\subsection{Observables}
Then, for each simulated star corresponding to a given population, we draw observables (position, velocity, metallicity, surface gravity) from plausible distributions for that population.  

\subsubsection{Positions and velocities for members}
\label{subsec:dfs}
As mentioned above, for dwarf members we randomly draw projected positions and l.o.s. velocities from the same phase-space distribution functions that we have previously used to generate mock data sets in previous studies (\citealtads{2011MNRAS.418.1526C,2015MNRAS.446.3002B}).  We have chosen four models in order to capture a range of behaviours corresponding to various plausible properties of dSph DM halos and the phase-space distributions of their stellar tracers.  One has a large ($\sim 100$ pc) `core' of constant central DM density; two have centrally `cusped' density profiles ($d\log\rho/d\log r=-1$ as $r\rightarrow 0$), as in halos produced in cosmological N-body simulations \citepads{2008MNRAS.391.1685S}; the fourth has a shallower cusp ($d\log\rho/d\log r=-0.6$ as $r\rightarrow 0$), mimicking a possible effect of stellar feedback in low-mass halos \citepads{2014ApJ...786...87B}.  One model has a stellar velocity dispersion profile that increases with radius; another's velocity dispersion profile decreases with radius, and two have velocity dispersion profiles that are approximately constant.  Among the four models, global stellar velocity dispersions range from $4.5$ km s$^{-1}$ to $11$ km s$^{-1}$.  We draw 1000 mock data sets for each model, giving a total of 4000 mock `ultrafaints'.

\subsubsection{Positions and velocities for contaminants}
In order to allow for varying degrees of overlap with the Galactic velocity distribution (held constant; see third bullet point below), we shift the mean l.o.s. velocity of dwarf members by a value that we draw randomly from the interval $-200\leq \langle v\rangle \leq +200$ km s$^{-1}$. 
 \begin{itemize}
   \item For stream and Galactic foreground stars, we draw projected positions from distributions that are uniform over the field subtended by dwarf members. 
   \item For stream stars, we draw l.o.s. velocities from a Gaussian distribution that has the same mean velocity as the dwarf members, but---in order to represent a more massive parent system---twice the velocity dispersion \citepads{2009MNRAS.397.1748B}. 
   \item For Galactic foreground stars, we draw l.o.s. velocities from the distribution that is expected to be observed along the l.o.s. to the Fornax dSph, based on the Galactic model of Besancon \citepads{2003A&A...409..523R}.
 \end{itemize}
All mock observables are scattered by $4$ km s$^{-1}$ in velocity in order to mimic realistic observational errors (e.g., \citealtads{2007ApJ...670..313S}).

\subsubsection{Metallicity and surface gravity}
For the member stars within a given realisation, we adopt a mean metallicity that is drawn from a distribution that is uniform between $-3.5\leq \langle [\rm Fe/H]\rangle \leq -1.5$.  For a given member star within that realisation, we then draw metallicity from a Gaussian distribution with the adopted mean and dispersion of $\sigma_{\rm [Fe/H]}=0.5$ dex, typical of observed ultrafaints \citepads{2010ApJS..191..352K}.   \begin{itemize}
  \item In all realisations, we draw metallicities for stream stars from a Gaussian distribution with mean $\langle \mathrm{[Fe/H]}\rangle = -1.5$ and dispersion $\sigma_{\mathrm{[Fe/H]}}=1$ dex. 
  \item For member and stream stars, we draw surface gravity from a Gaussian distribution with mean $\langle \log_{10} g\rangle = 1.5$ (cgs units) and dispersion $\sigma_{\log g}=1$ dex. 
  \item For Galactic foreground stars, we draw metallicities from a Gaussian distribution with mean $-0.5$ dex and dispersion 1.5 dex, and surface gravities from a Gaussian distribution with mean $\langle \log_{10} g\rangle = 4$ and dispersion 1 dex.
\end{itemize}
All mock observables are scattered by $0.2$ dex in metallicity and $0.2$ dex in surface gravity in order to mimic realistic observational errors (e.g., \citealtads{2010ApJS..191..352K}).

\subsubsection{Membership probabilities}
For each star in a given mock data set, we estimate probability of membership using the expectation-maximisation (EM) algorithm described by \citetads{2009AJ....137.3109W}.  The EM algorithm iteratively uses estimates of parameters (e.g., means and dispersions) that specify distributions of observables (e.g., velocity, metallicity, surface gravity) in order to evaluate membership probabilities for individual stars, then uses those probabilities to update estimates of model parameters. 

Because we use a single Gaussian distribution to represent the velocity distribution of member stars, our implementation of the EM algorithm  effectively assumes that the velocity dispersion profile is flat, introducing potential biases when the dSph has a radially-variable velocity dispersion profile.   Despite our use of four different dynamical models whose velocity dispersion profiles range from decreasing to flat to increasing (Section \ref{subsec:dfs}), we did not find any systematic errors that were specific to one particular kind of model.  We conclude, therefore, that our current study specifically examines the effect of sample contamination and does not depend strongly on the nature of the underlying dynamical model.

\section{$J$-factors for contaminated mock data}
\label{sec:mock_tests}

\subsection{Uncertain membership status and contamination}
\begin{figure}
\begin{center}
\includegraphics[width=\columnwidth]{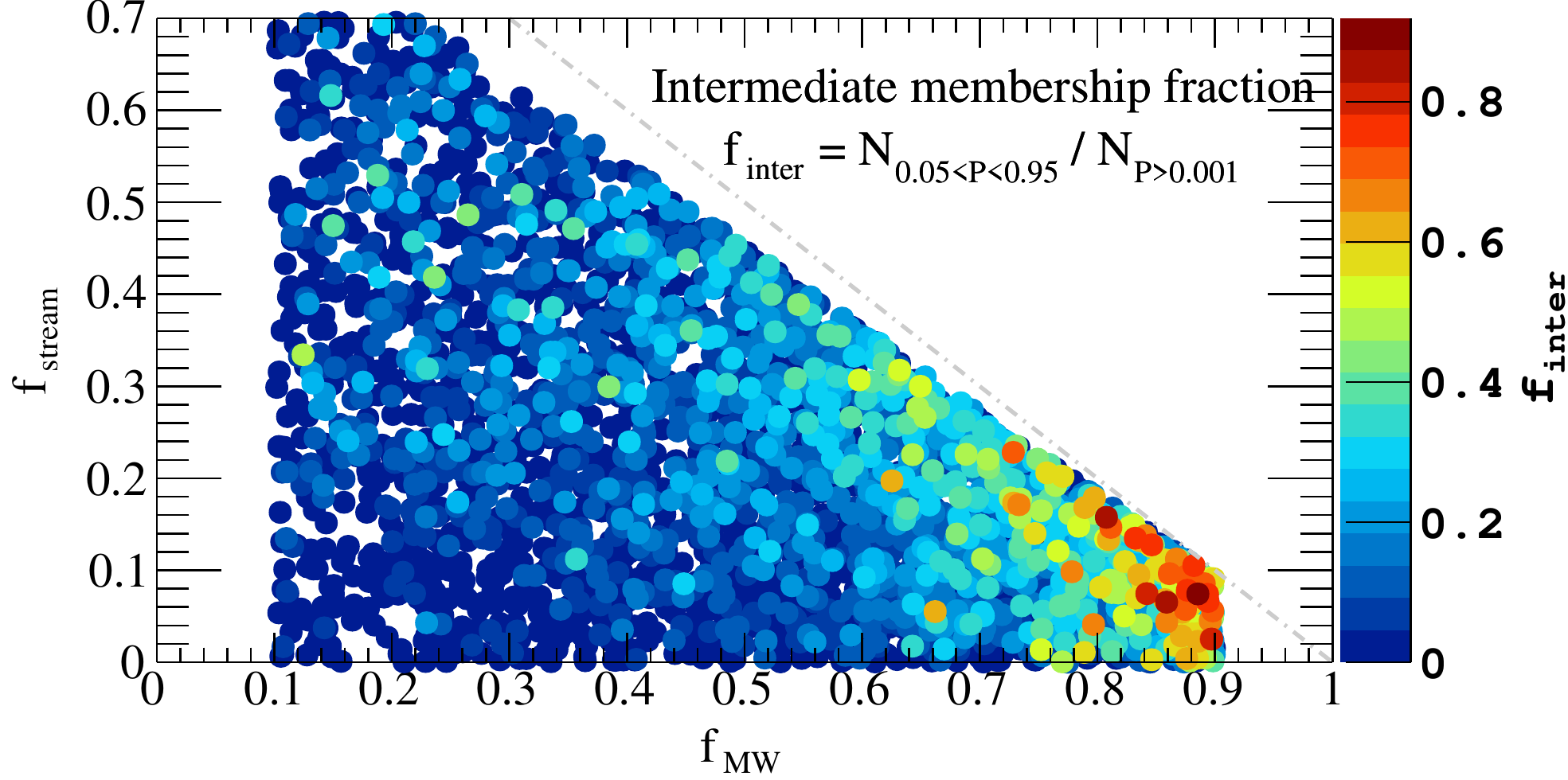}
\includegraphics[width=\columnwidth]{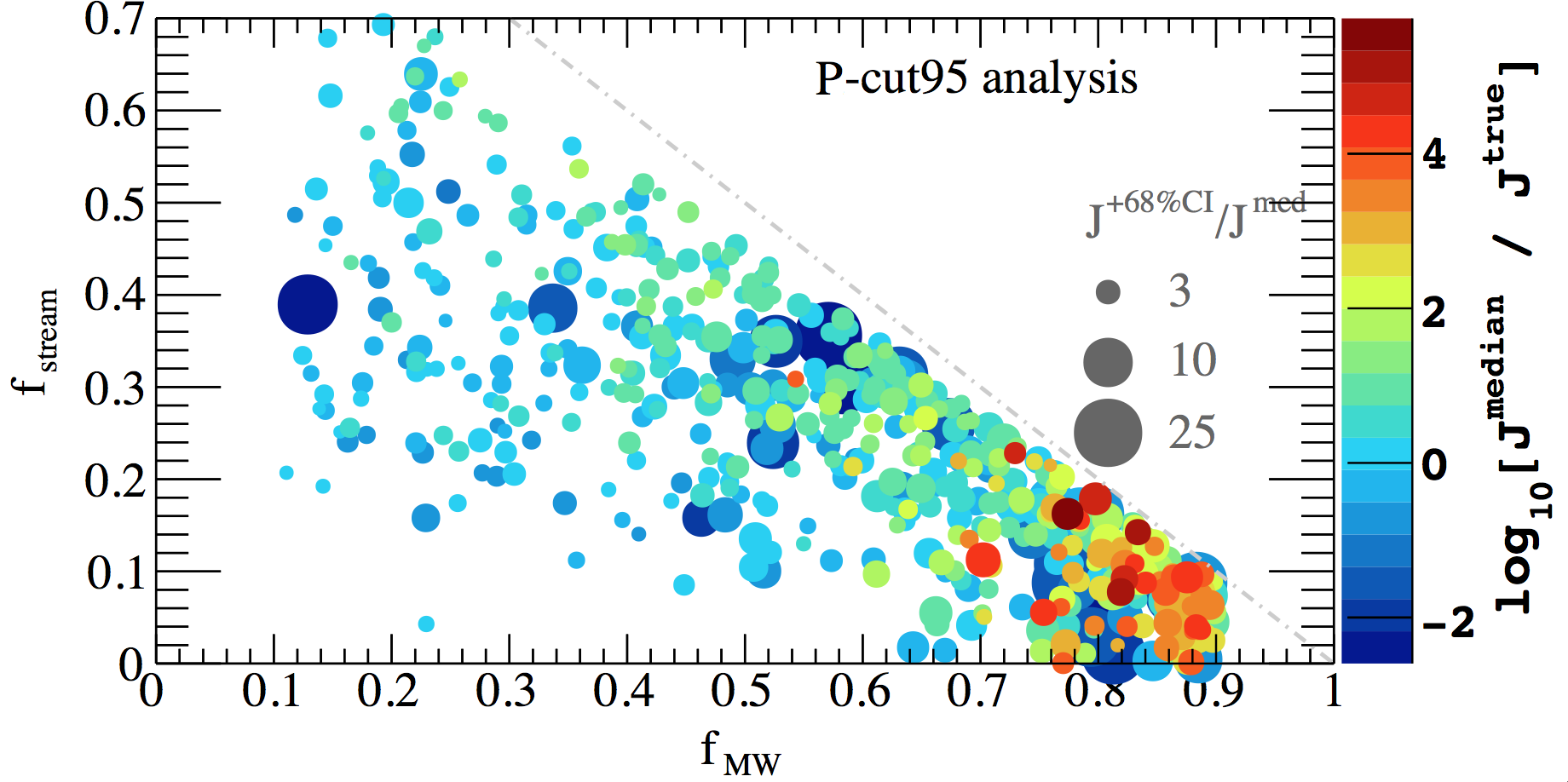}
\caption{\textit{Top}: fraction $f_{\rm inter}$ of plausible members (membership probability $P > 10^{-3}$) whose membership status is ambiguous ($0.05 < P < 0.95$) for our 4000 mock data sets, as a function of the fractions of the sample contributed by contamination from the Milky Way ($x$ axis) and stream ($y$ axis). \textit{Bottom}: accuracy of estimated $J$-factors (MCMC/cut-95\% analysis, $\alpha_{\rm int} = \alpha_c$), for the 545 mock dSphs with $f_{\rm inter} > 0.1$.  The colour scale shows deviation of the estimated $J$-factor from the true value.  The sizes of the symbols correspond to the size of the 68\% CI on the estimated $J$-factor.}
\label{fig:fig1}
\end{center}
\end{figure}

We first look at the impact of contamination on the estimation of the membership probabilities with the EM algorithm.  Based on their study of 27 mock data sets of different levels of contamination from simulated foreground populations, \citetads{2009AJ....137.3109W} conclude that the algorithm reliably separates contaminants from dSph members, except when samples are small ($N_{\rm stars} \la 30$) and suffer heavy contamination (i.e., when there is a small number of members buried among a large sample of contaminants).

Here, using our 4000 `ultrafaint'-like mock dSphs, we focus on the impact of stars that show ambiguous membership status, i.e., have intermediate membership probabilities.  We define a new diagnostic, $f_{\rm inter}$, as the fraction of plausible members that have ambiguous status.  Specifically, $f_{\rm inter}$ is the ratio of the number of stars with $0.05< P < 0.95$ to the number of stars with $P > 10^{-3}$.  This quantity is useful because it can be computed for real data sets. For our 4000 mock data sets, Figure \ref{fig:fig1} shows how $f_{\rm inter}$ depends on intrinsic levels of contamination by simulated Milky Way and stream stars, where the fractions of stream and Milky Way stars in a given mock data set are given by $f_{\rm MW}=1-f_1$ and $f_{\rm stream}=f_1(1-f_2)$, respectively (see Section \ref{subsec:simpops}).  

Strong levels of contamination, particularly from the Milky Way, tend to result in large fractions of stars having ambiguous membership status. Therefore, the quantity $f_{\rm inter}$ can serve as a proxy for contamination levels in real dSphs, and will be discussed for the cases of the `ultrafaint' dSphs Ret~II and Seg~I in Section~\ref{sec:seg1}.  Henceforth, in order to focus on situations where sample contamination is important, we consider only the 545 mock data sets for which $f_{\rm inter}>10\%$ (see bottom panel of Fig.~\ref{fig:fig1}). 

\subsection{Impact of contamination on $J$-factors}
\label{subsec:impactonj}
\paragraph*{Four different procedures}
For each of the 545 mock data sets with $f_{\rm inter}> 10\%$, we perform both the MCMC and bootstrap analyses described in Sections \ref{subsec:MCMC} and \ref{subsec:bootstrap}.  For both analyses we handle the estimated membership probabilities in two different ways, both of which have been employed by various authors in analyses of real dSph data.  In the first case, we use all data points and weight each by its probability of membership according to Eq. (\ref{eq:likelihood}); henceforth, we refer to this procedure as `$P$-weighted'.  In the second, we discard all observations for which the probability of membership fails to exceed a fiducial threshold of $P> P_{\rm thresh}$, and we give equal weight to all observations that exceed this threshold.  In the second case, we still use Eq. (\ref{eq:likelihood}) as the log-likelihood, but only after reassigning membership probabilities to $P=0$ for all stars originally below the threshold and to $P=1$ for all stars originally above it.  Following several published analyses (e.g., \citealtads{2011MNRAS.418.1526C}), we adopt $P_{\rm thresh}=0.95$, and henceforth refer to this procedure as `cut-95\%'. Note that the membership probabilities themselves are likely to carry large uncertainties, but the latter are not estimated by the  
different reconstruction algorithms (e.g., the EM algorithm). We are therefore left to deal with their point estimates.

Considering the MCMC and bootstrap analyses and both scenarios for handling membership, we obtain four independent estimations of the $J$-factor for each mock `ultrafaint', which we compute here at $\alpha_{\rm int} = \alpha_c = 2\times r_h/d$ \citepads{2011ApJ...733L..46W,2011MNRAS.418.1526C}, with $r_h$ the half-light radius and $d$ the distance to the mock dSph\footnote{Note that our conclusions do not depend on the distance to the mock dSph.}.  We then compare the estimated $J$-factors to the `true' values that we calculate directly from the DM density profiles that we used to generate the mock data.  In what follows, all statistical uncertainties are $1\sigma$, representing 68\% CIs for the MCMC analyses and $\sigma_{\rm bootstrap}$ for the bootstrap analyses.

\paragraph*{$J$-factor overestimation}
The bottom panel of Figure \ref{fig:fig1} depicts accuracy of the $J$-factor estimates in the $f_{\rm MW}-f_{\rm stream}$ plane for the 545 mock dSphs.  These particular results correspond to the cut-95\% procedure with MCMC parameter estimation (see below for a comparison of results from all four procedures).  Dark blue symbols correspond to cases where $J$ is underestimated; these are rare and tend to have relatively large CIs.  Colours ranging from green to red represent cases where $J$ is overestimated by factors as large as $\sim 10^5$.  These cases tend to occur when samples have a large degree of contamination, near the line $f_{\rm MW}+f_{\rm stream}=1$.  The largest overestimates are driven by foreground contaminant stars that are mistaken for members, with the EM algorithm returning $P\ga 0.95$.  The velocity dispersion of the Milky Way stars is significantly larger than the one from the member stars, and the mis-classification of even a few such contaminants can mimic large DM densities and yield large $J$-factors.  Alarmingly, even when $J$ is severely overestimated, the CIs can be relatively small.  Since real dSphs with large and tightly-constrained $J$-factors receive the greatest weight in subsequent inferences of cross-sections for DM annihilation, it is important to recognise when such estimates may be biased by sample contamination.  

\begin{figure}
\begin{center}
\includegraphics[width=\columnwidth]{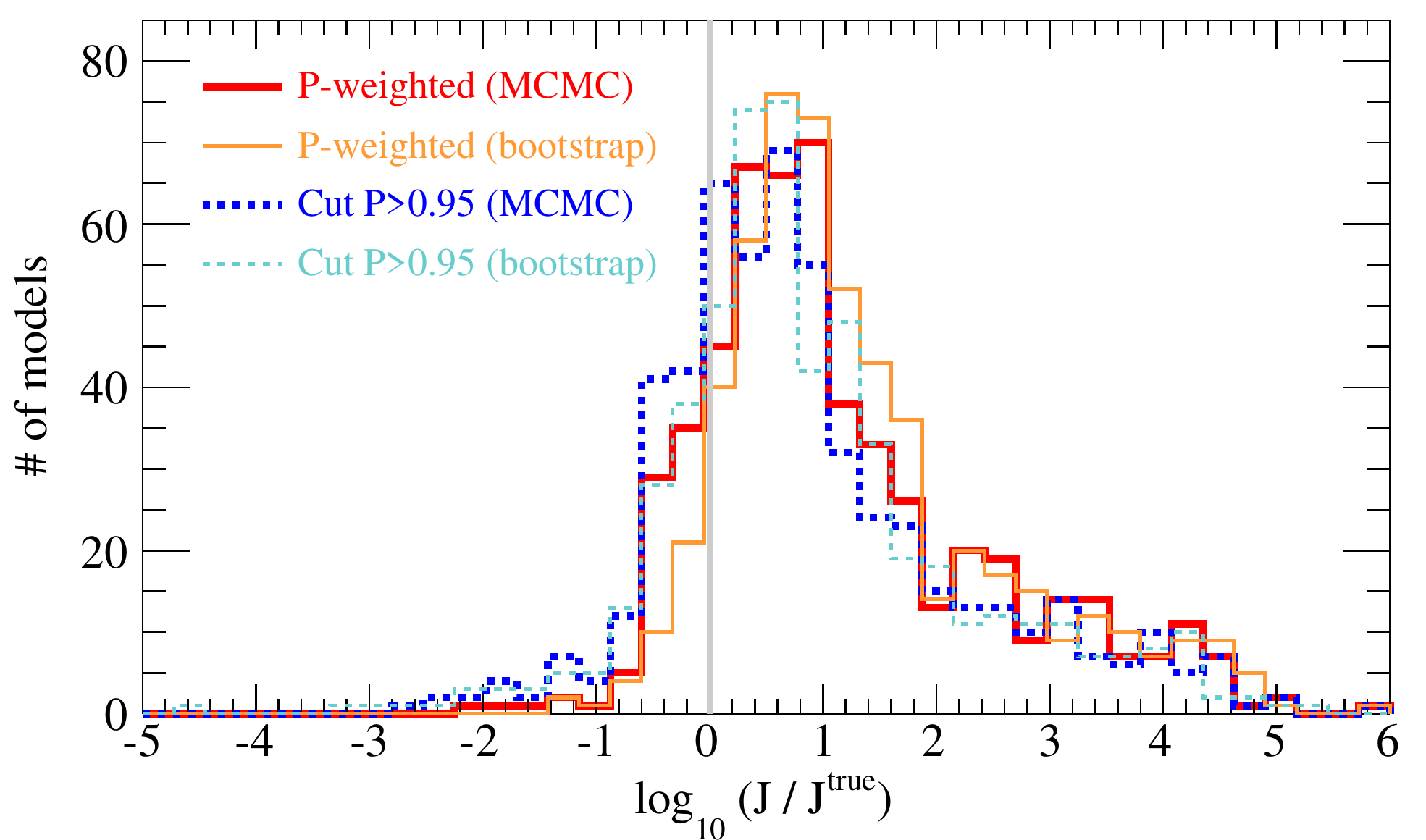}
\includegraphics[width=\columnwidth]{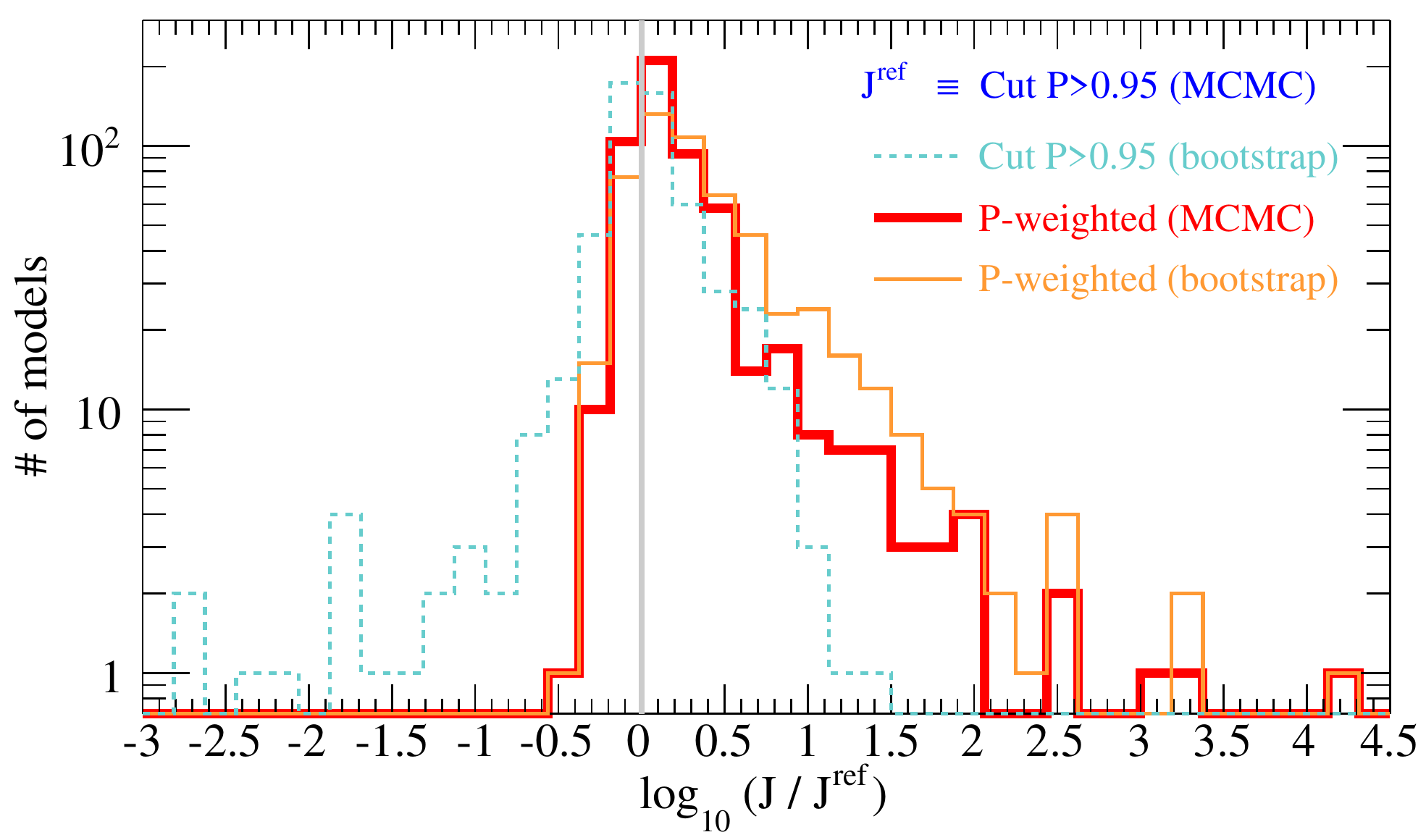}
\caption{\textit{Top}: distributions of $\log_{10}(J/J_{\rm true})$ at $\alpha_{\rm int} = \alpha_c$, for the 545 mock dSphs, each analysed with four different procedures that differ according to method of parameter estimation (MCMC vs. bootstrap) and treatment of membership probabilities (weighting of individual data points vs. discarding low-probability members).  All procedures yield long tails corresponding to overestimates of $J$. \textit{Bottom}: distributions of relative differences between the four procedures; the `reference' procedure is taken to be the case in which we estimate parameters via MCMC and discard stars for which estimated probabilities of membership are smaller than $0.95$. 
}
\label{fig:fig2}
\end{center}
\end{figure}

The top panel of Figure \ref{fig:fig2} shows distributions of $\log_{10}(J/J_{\rm true})$ at $\alpha_{\rm int} = \alpha_c$ for the 545 mock data sets with $f_{\rm inter}>10\%$, and for the four different analysis procedures. All four yield similar distributions, with long tails extending toward overestimated $J$, reaching values as large as $\log_{10}(J/J_{\rm true})\sim 5$. However, the most extreme cases of overestimation are due to strong levels of Milky Way contamination, which give very large velocity dispersions that would be easily identified when analysing real dSphs. If we select for instance the models for which the mean stellar velocity dispersion $\bar{\sigma_p}$ is smaller than $20$ km~s$^{-1}$, the maximum overestimation reduces to $\sim$ 3 orders of magnitude. The distribution remains however skewed toward overestimated values because of the contamination from both Milky Way and stream stars, which is not the case with contamination-free mock data \citepads{2015MNRAS.446.3002B}.

\paragraph*{Comparing results from different procedures}
In order to examine sensitivity to choice of procedure, we systematically compare results from the four different procedures that we applied to each mock data set. For ease of comparison, we refer the result from each procedure to the one obtained from our MCMC analysis of the cut-95\% procedure that was used to construct Figure \ref{fig:fig1}. The bottom panel of Figure \ref{fig:fig2} show distributions of $\log_{10}(J/J^{\rm ref})$ obtained for our 545 samples.

For most of the models, the four procedures give similar results, such that distributions of $\log_{10}(J/J^{\rm ref})$ peak near zero.  However, when procedures give different estimates of $J$, we tend to get larger values when using the membership probabilities as weights.  This is true for both MCMC and bootstrap methods (red and orange distributions respectively).  The reason is that lower-probability members tend to populate tails of the members' velocity distribution. Analyses that give non-zero weight to these stars will necessarily operate on broader velocity distributions and infer larger DM masses.  

\begin{figure}
\begin{center}
\includegraphics[width=\columnwidth]{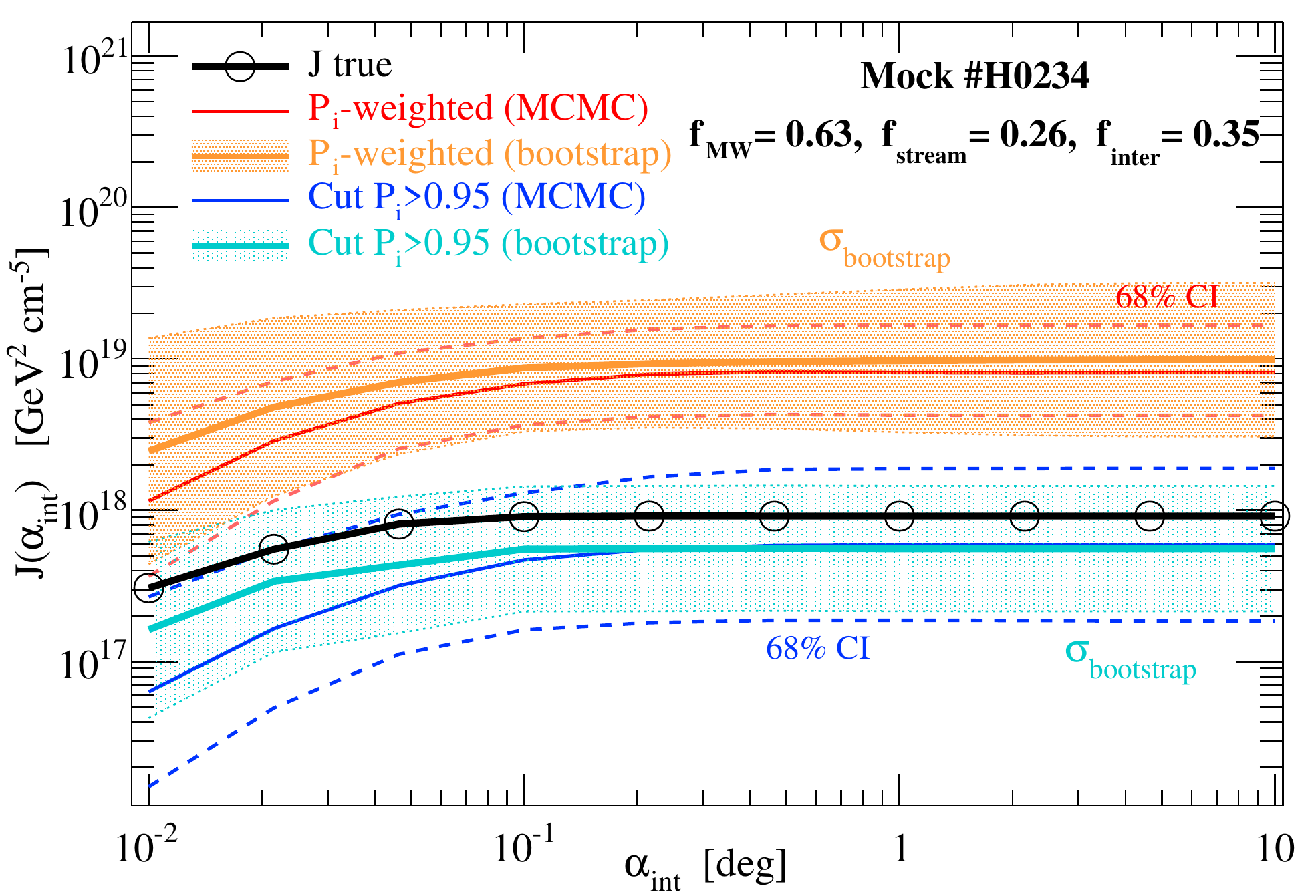}
\includegraphics[width=\columnwidth]{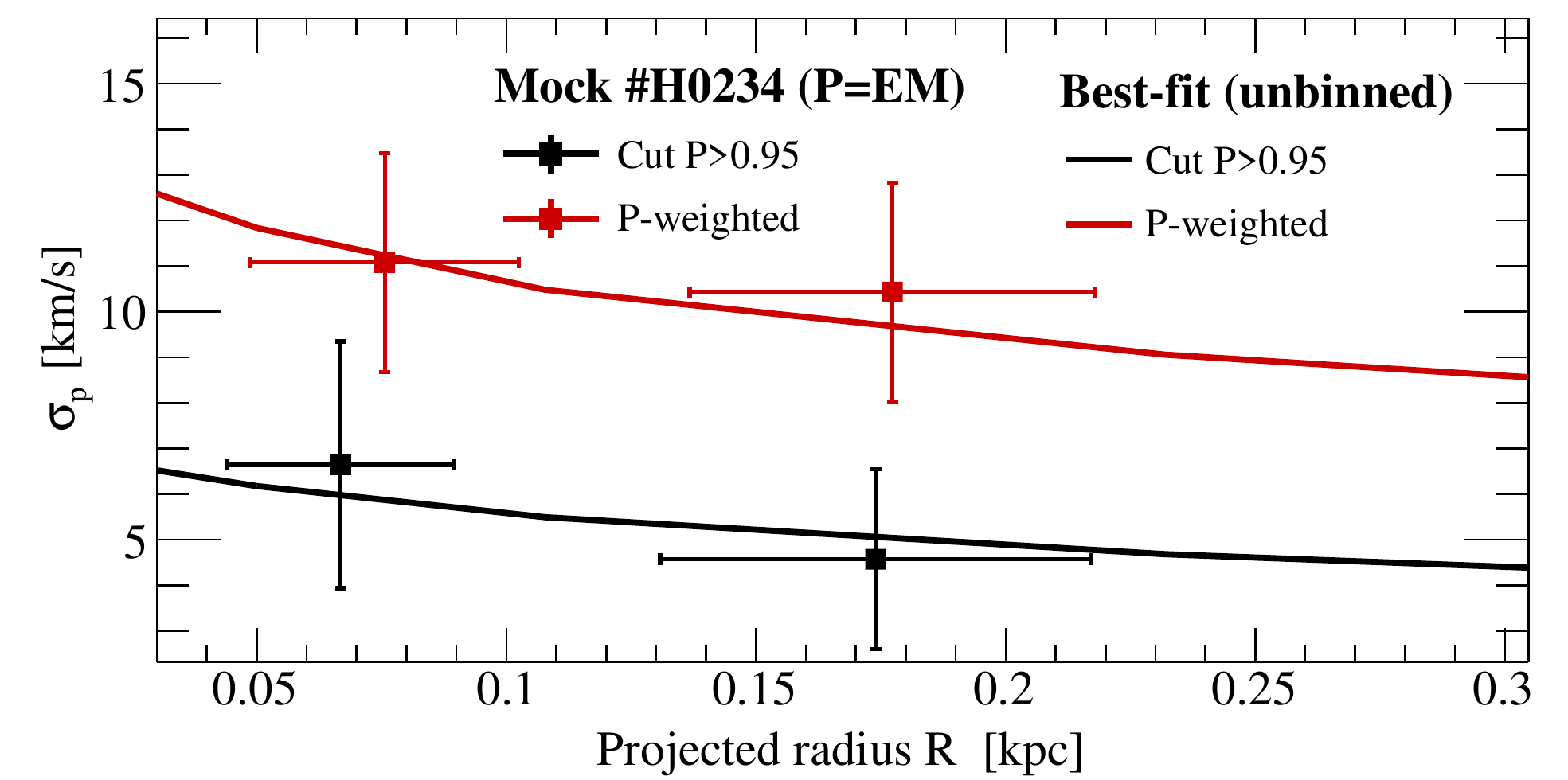}
\includegraphics[width=\columnwidth]{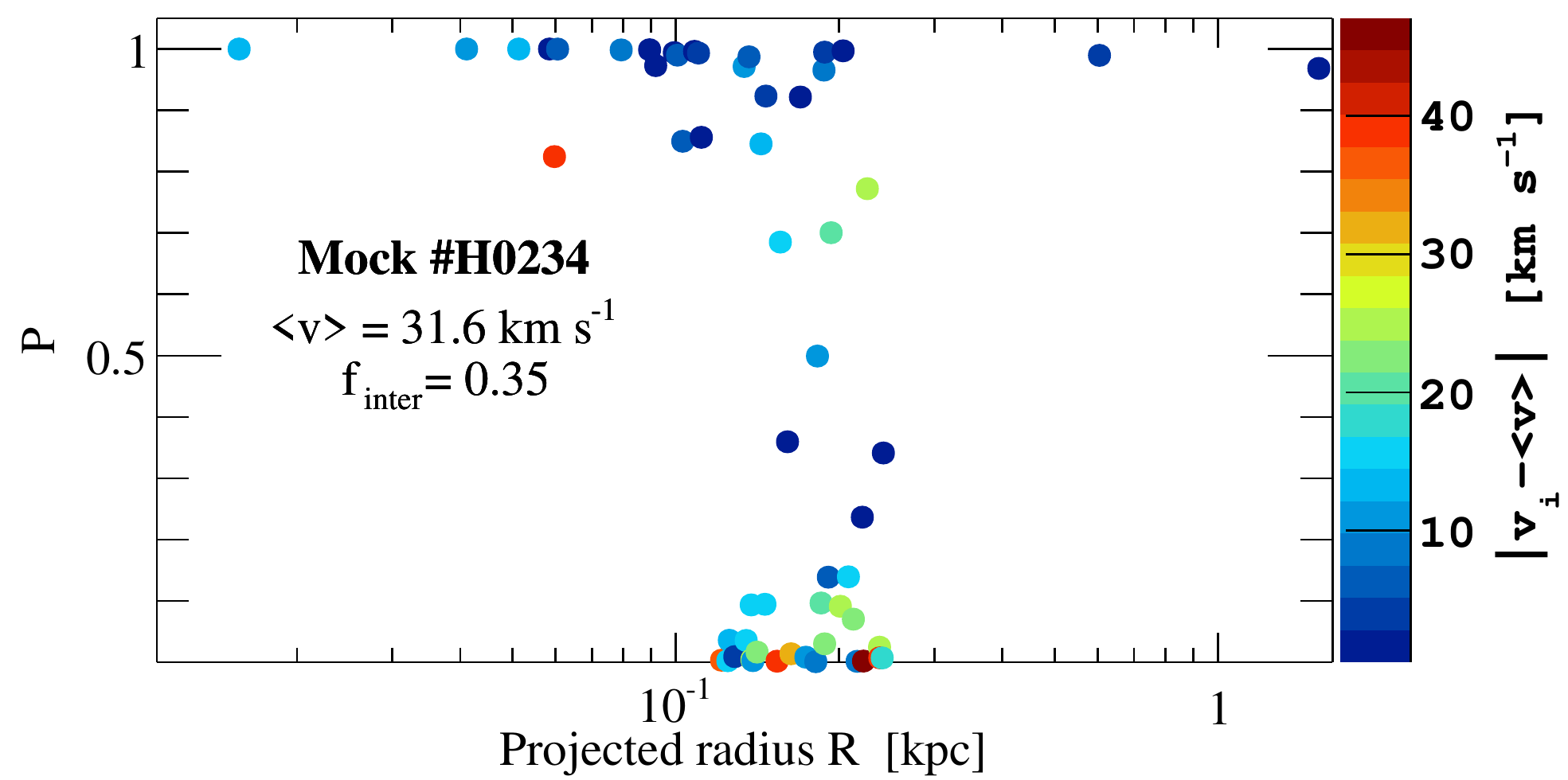}
\caption{\textit{Top: } $J$-factor as a function of integration angle for the mock ultrafaint dSph H0234. The true $J$-factor, in solid black, is compared to estimates obtained with four different analyses that differ based on choice of procedure for parameter estimation (MCMC vs. bootstrap) and treatment of membership probabilities (weighting of all observations by membership probability vs. discarding all observations with $P \leq 0.95$). Filled regions and dashed curves correspond to $1\sigma$ uncertainties obtained via bootstrap and MCMC procedures, respectively. \textit{Middle:} line-of-sight velocity dispersion profile for H0234, estimated either using membership probabilities as weights (red) or considering only observations for which probability of membership is $>0.95$ (black). The best-fit models from the \textit{unbinned} MCMC analyses are shown for illustration purpose. \textit{Bottom:} for the same mock dSph, deviation from the mean velocity, $|v-\langle v\rangle |$ (blue to red colours), as a function of membership probability, $P$ (only stars with $P>10^{-3}$ are plotted), and projected radius, $R$.}
\label{fig:fig3}
\end{center}
\end{figure}

Figure \ref{fig:fig3} provides a concrete example, showing results from our analysis of one particular mock data set.  The bottom panel shows how velocity and membership probability depend on projected distance from the dSph center.  The fraction of plausible members with ambiguous status is $f_{\rm inter}=0.35$. Depending on how we treat these stars, the velocity dispersion profile changes dramatically: it is smaller by a factor $\sim 2$ when removing the stars with $P < 0.95$ (middle panel).\footnote{We also show the best-fit models obtained in each case from the MCMC analysis. As we employ an \textit{unbinned} analysis, these fits are shown only for illustration purpose.} The top panel shows how our estimate of $J$ depends strongly on how we treat these stars.  The $P$-weighted procedure gives non-zero weight to the ambiguous stars, which in this case are non-members.  As a result, the $P$-weighted procedure overestimates $J$ by two orders of magnitude and, despite this inaccuracy, has relatively small CIs (dashed red curves and filled orange zone, for MCMC and bootstrap analyses respectively).  In contrast, the cut-95\% procedure gives zero weight to ambiguous cases and recovers the true value of $J$ within its broader CIs.  

\paragraph*{Lessons from analysis of mock data}
Our experimentation with mock data sets teaches us that when stellar-kinematic samples are small and heavily contaminated, $J$-factors tend to be overestimated---sometimes by orders of magnitude---as the result of mis-classification of even a few non-member stars. This conclusion stands, even if a more sophisticated method is used to simultaneously estimate the membership status and constrain the DM profile. This method is presented in App.~\ref{subsubsec:mixture}. Being more computationnaly demanding, it could not be tested extensively on the mock data, and such detailed analysis is beyond the scope of this paper. It was however found to give similar error bars despite propagating correctly the membership uncertainties to the DM profile estimation --- the uncertainties remain dominated by the sample size --- and was not able to better retrieve the correct $J$-factor. For the pathological H0234 mock model discussed above, it performs even worse than the cruder standard analysis, overshooting the correct $J$-factor by several orders of magnitude. For these reasons, the two-step approach (extracting membership probabilities and then using them in the Jeans analysis) seems adequate for identifying cases where contamination is likely to be particularly problematic.

In the context of this approach, cases of severe overestimation tend to exhibit characteristics that can be identified in real data sets.  Warning signs include:

\begin{itemize}
\item relatively large fractions of stars with ambiguous membership status, as quantified by $f_{\rm inter}$;
\item sensitivity of $J$-factor estimates to the exact procedure chosen for addressing membership (e.g., using membership probabilities for all stars vs. discarding observations below some threshold).
\end{itemize}

\section{Segue~I and Reticulum~II}
\label{sec:seg1}

Having gained intuition from our systematic analysis of mock data sets, we now consider real stellar-kinematic data sets that are publicly available for two `ultrafaint' dSphs: Seg~I \citepads{2011ApJ...733...46S} and Ret~II \citepads{2015ApJ...808..108W,2015ApJ...808...95S,2015ApJ...811...62K}.  

\subsection{Reticulum~II}
We first focus on the kinematic sample of Ret~II.  Following \citetads{2015ApJ...808L..36B}, we use the kinematic data of \citetads{2015ApJ...808..108W}, which contains estimates of velocities, effective temperatures, surface gravities and metallicities for 37 red giant candidates along the l.o.s. to Ret II\footnote{We confirm that we obtain the same results if we use instead the stellar-kinematic samples of \citetads{2015ApJ...808...95S} or \citetads{2015ApJ...811...62K}.}. We quantify membership probabilities using the EM algorithm of \citetads{2009AJ....137.3109W}. The sum of membership probabilities is $\approx 16.7$, where 16 stars have $P >0.99$ and another has $P \approx 0.7$.  

\begin{figure}
\begin{center}
\includegraphics[width=\columnwidth]{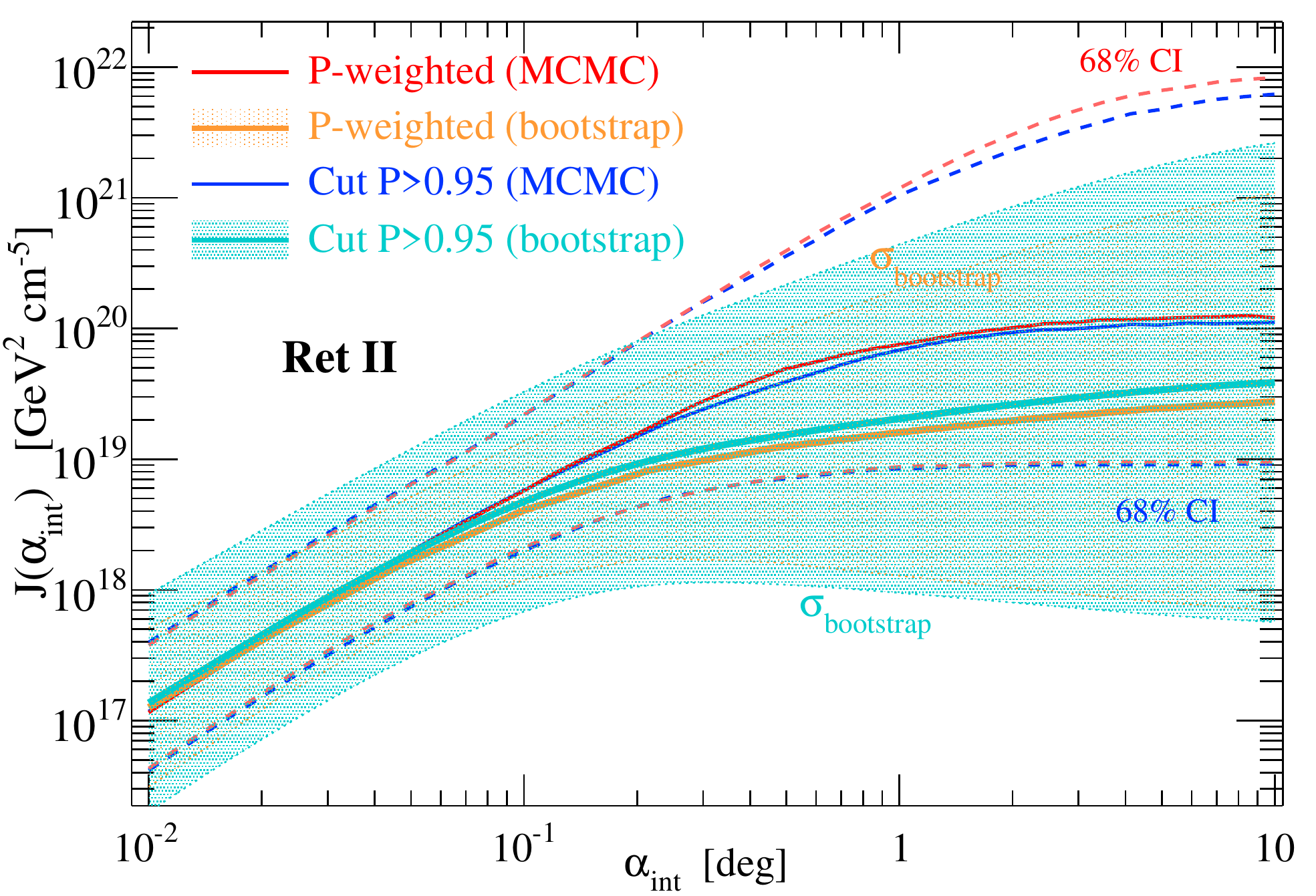}
\includegraphics[width=\columnwidth]{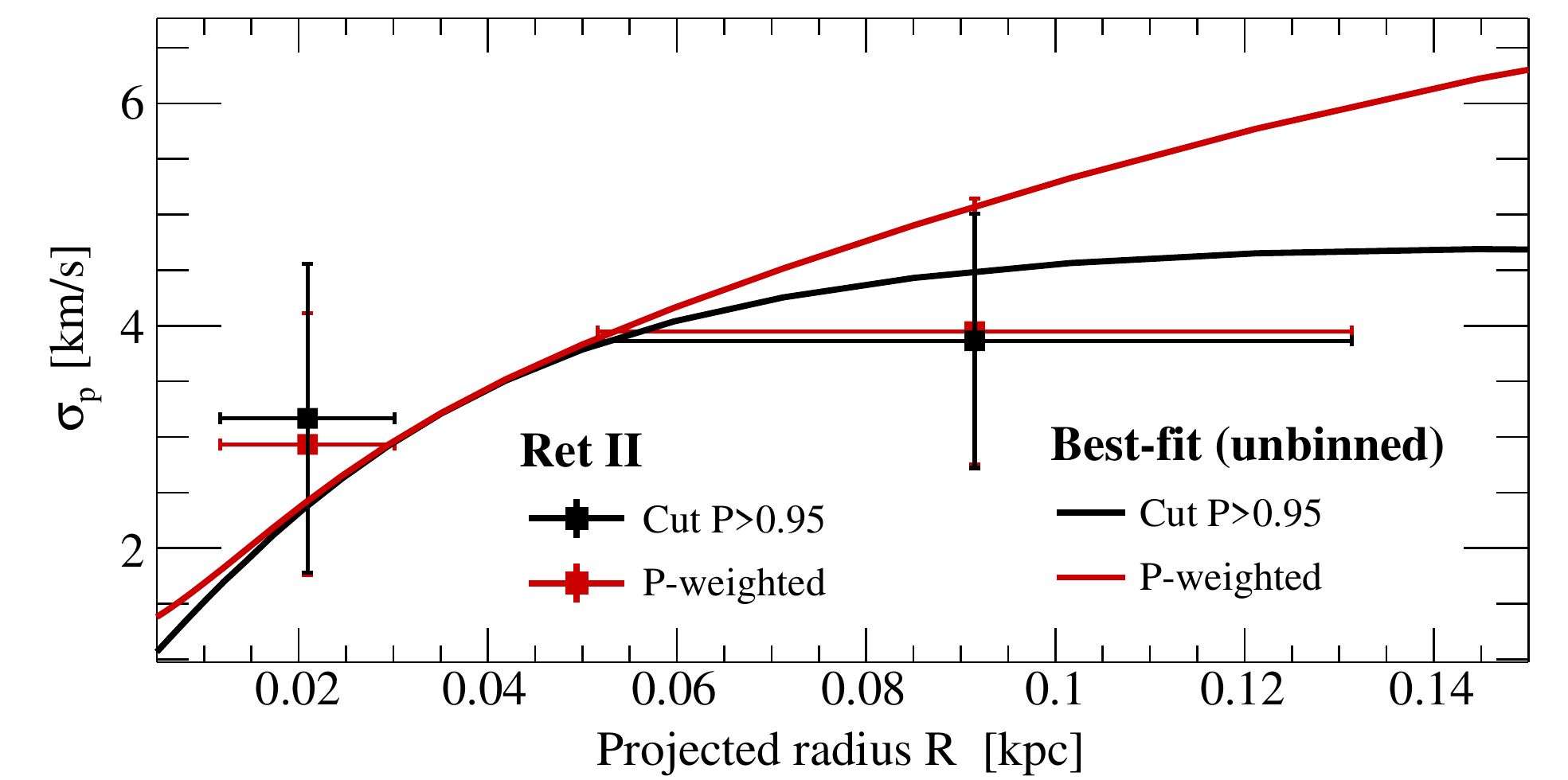}
\includegraphics[width=\columnwidth]{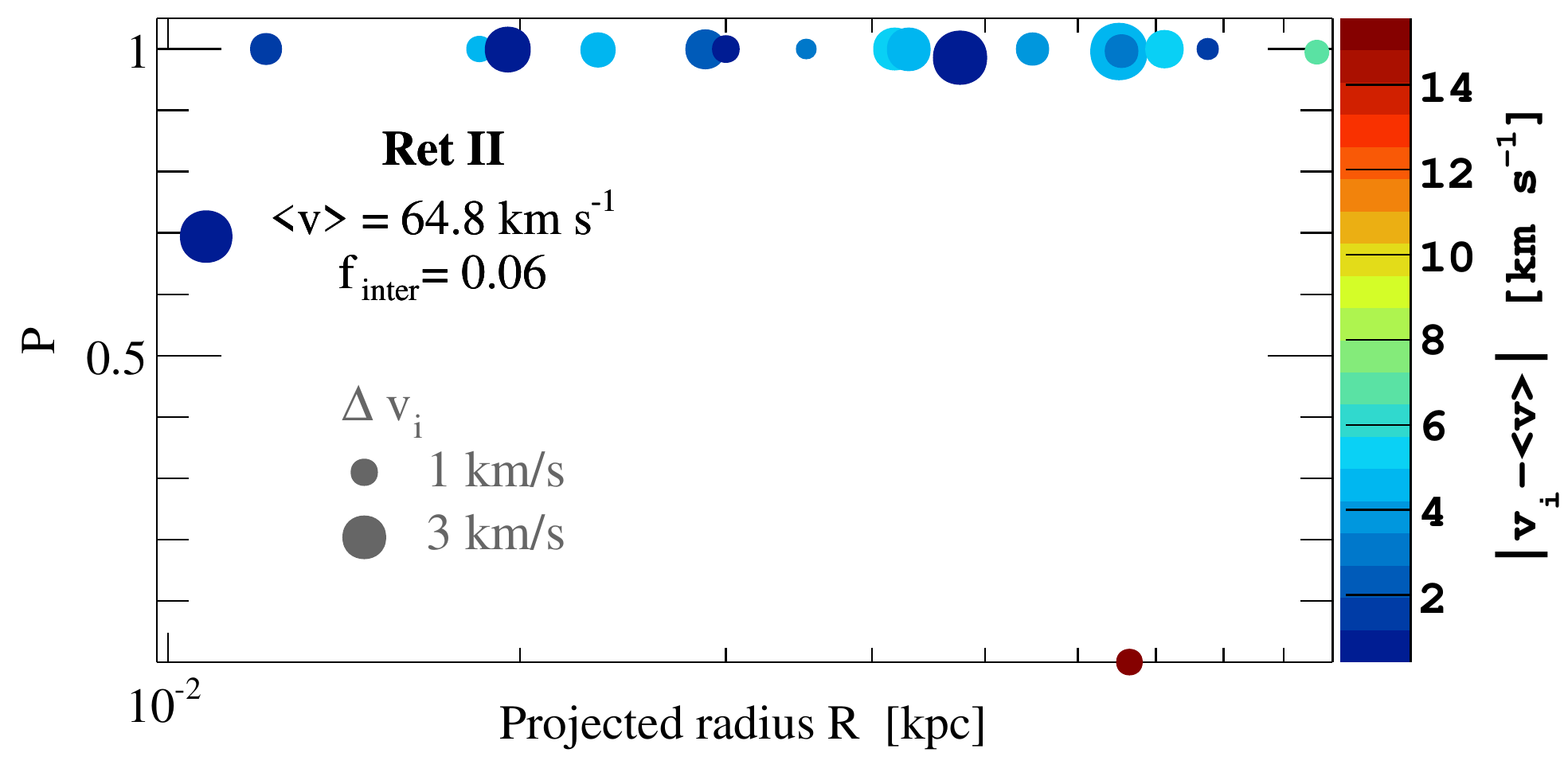}
\caption{Same as Figure \ref{fig:fig3}, except for the real `ultrafaint' dwarf Ret~II, using the stellar-kinematic data of \citeads{2015ApJ...808..108W}). Only one star has ambiguous membership probability ($P\approx 0.7$), and all four procedures for estimating $J$ give consistent results, indicating robustness against contamination of the stellar-kinematic sample. }
\label{fig:fig4}
\end{center}
\vspace{-0.3cm}
\end{figure}

The bottom panel of Figure \ref{fig:fig4} shows deviations from the mean velocity, $|v-\langle v \rangle|$ (blue to red colours), as a function of membership probability $P$ and projected radius $R$, for stars with $P>10^{-3}$. Only the star with $P\approx 0.7$ has ambiguous membership probability (due to its relatively large metallicity), and no obvious velocity outlier is visible. The velocity dispersion profiles obtained either by weighting the contribution of each star by its value of $P$, or by selecting only the stars with $P \leq 0.95$, are very similar (middle panel of Figure \ref{fig:fig4}). Applying our four different procedures gives confidence that our estimate of $J$ for Ret II is robust against contamination, as the $P$-weighted and cut-95\% analyses show good agreement for both MCMC and bootstrap analyses (top panel of Figure \ref{fig:fig4}).

\subsection{Segue~I}
\label{subsec:seg1}
For Seg~I, we analyse the stellar kinematic data set of \citetads[][`S11' hereafter]{2011ApJ...733...46S}, who measure velocity and equivalent width of the calcium-triplet absorption feature for 393 stars in the direction of Seg~I.  For the $\sim 180$ of these stars that have colours and magnitudes consistent with Seg~I membership, S11 estimate two different membership probabilities, using either the EM algorithm or a Bayesian technique described by \citetads{2011ApJ...738...55M}.  For straightforward comparison with our analyses of mock data sets, we first focus on the membership probabilities obtained from the EM algorithm.  Following S11, we remove from the sample two photometric variable stars (one of which S11 identify as a velocity variable), three additional stars that S11 identify as velocity variables, and one additional velocity outlier that S11 identify as having a strong effect on estimates of velocity dispersion.  

\begin{figure}
\begin{center}
\includegraphics[width=\columnwidth]{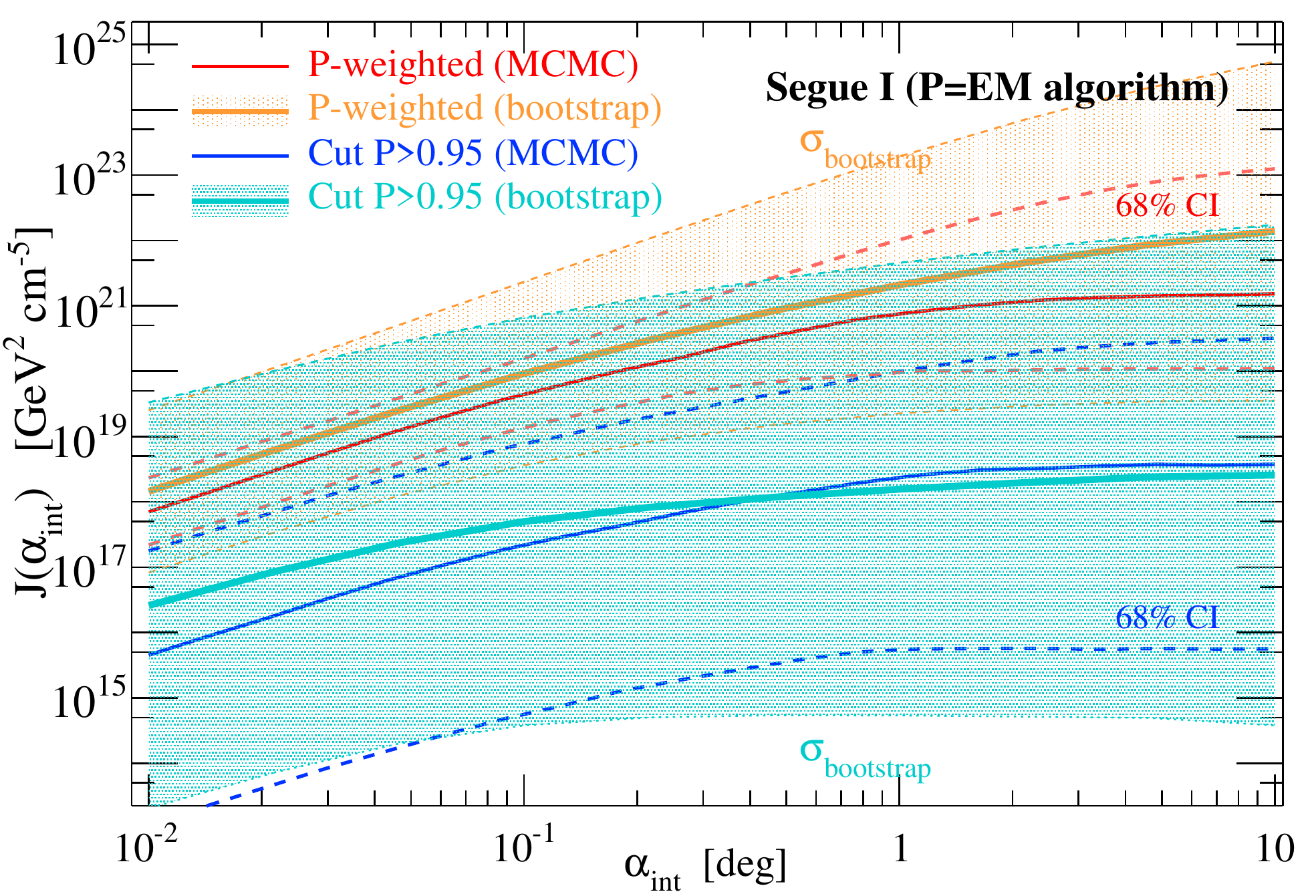}
\includegraphics[width=\columnwidth]{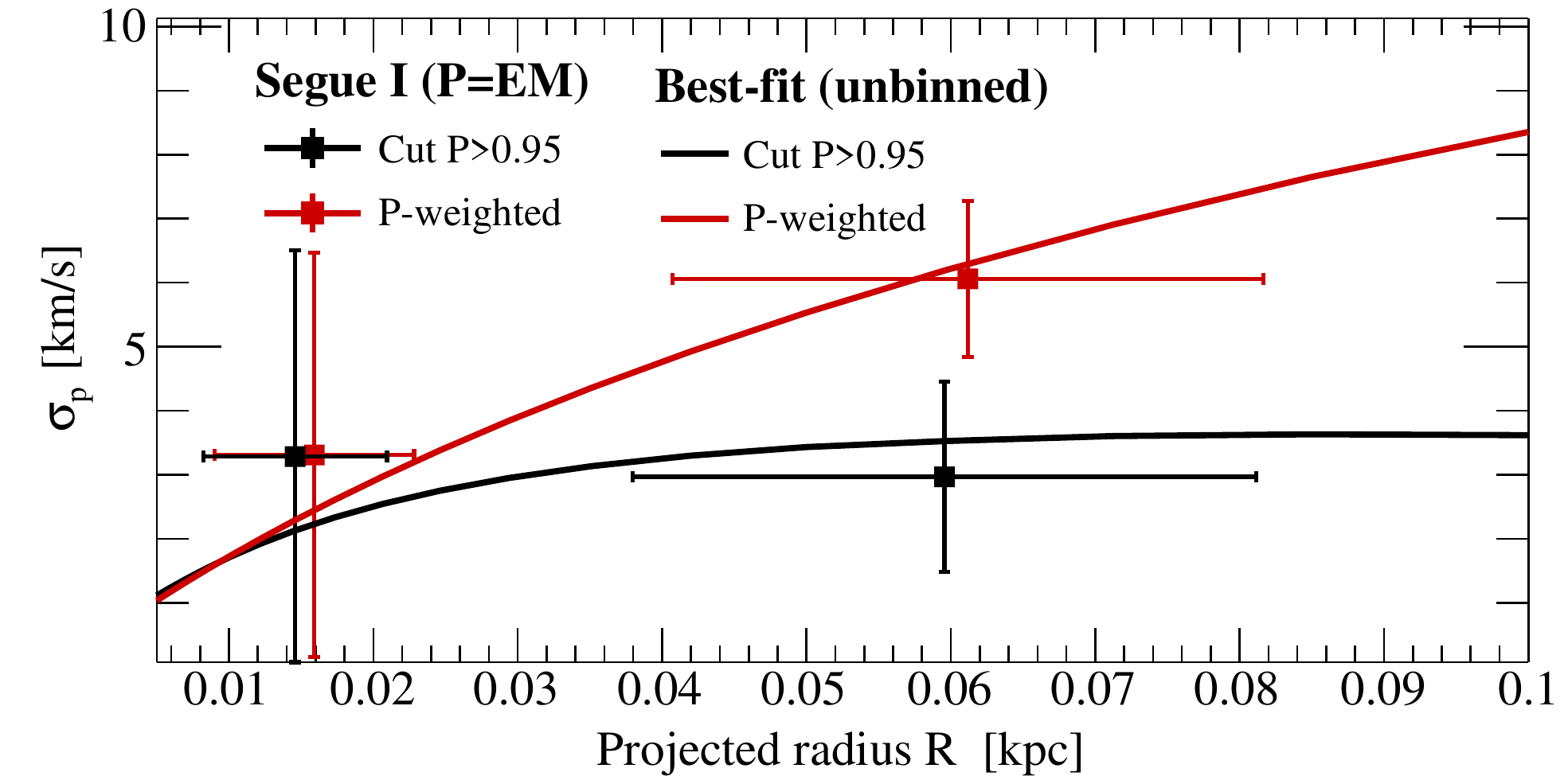}
\includegraphics[width=\columnwidth]{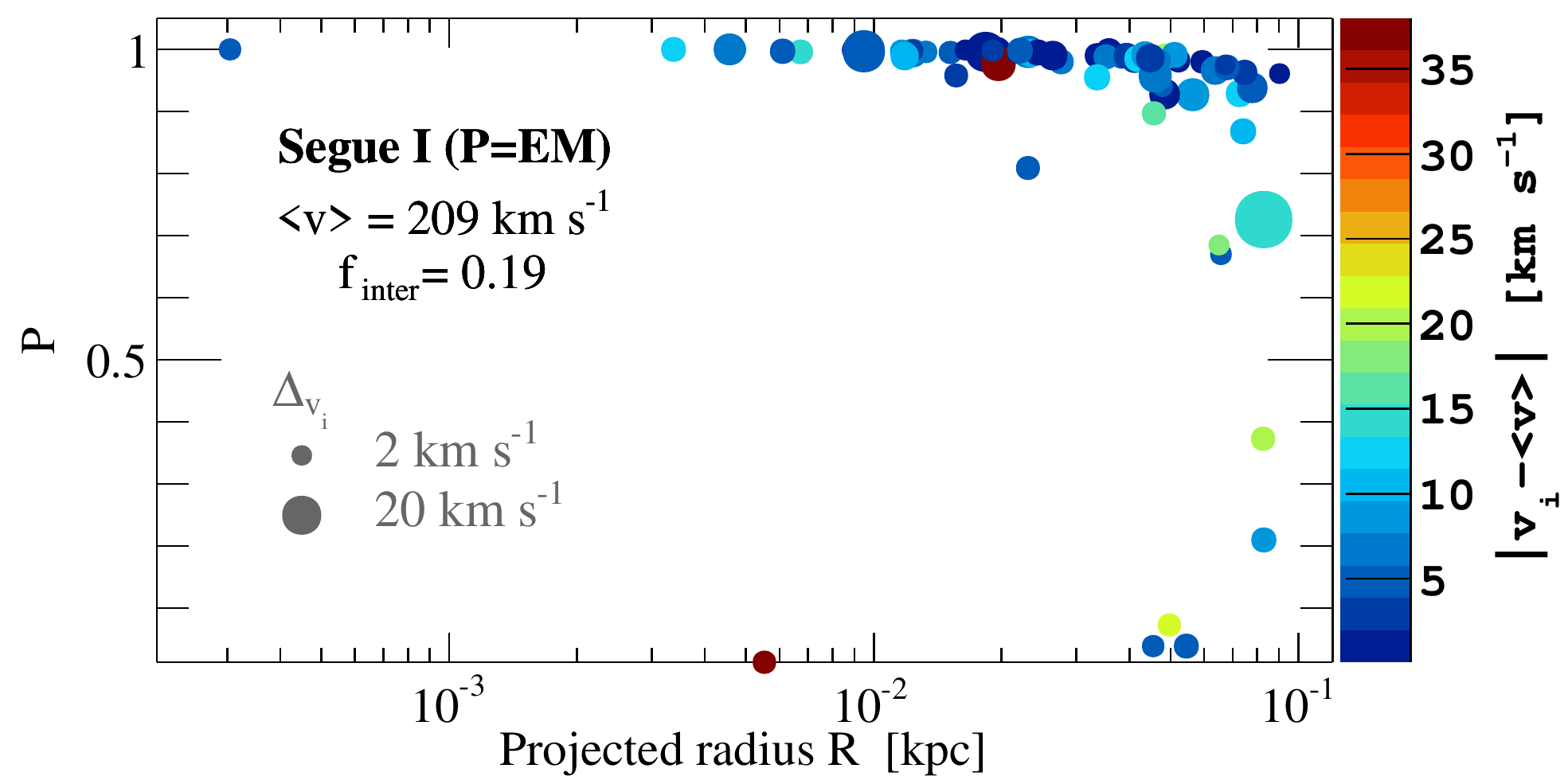}
\caption{Same as Figure \ref{fig:fig4}, but for Seg~I, using the stellar-kinematic sample of \citetads{2011ApJ...733...46S}.}
\label{fig:fig5}
\end{center}
\vspace{-0.3cm}
\end{figure}

Even after removing these stars, the remaining sample for Seg~I shares characteristics with the mock data sets for which we found estimates of $J$ to be unreliable.  The fraction of stars with ambiguous membership probability is $f_{\rm inter} = 0.19$ (bottom panel of Figure \ref{fig:fig5}).  Moreover, for both MCMC and bootstrap procedures for parameter estimation, our estimates of $J$ for Seg~I depend strongly on whether we use the cut-95\% or the $P$-weighted procedure to handle membership probabilities (top panel of Figure \ref{fig:fig5}). For example, using the $P$-weighted procedure, we obtain a $J$-factor at $\alpha_{\rm int} = 1^{\circ}$ that is more than two orders of magnitude larger than the result from the cut-95\% procedure.  This behaviour is strikingly similar to that of the mock data set illustrated in Figure \ref{fig:fig3}, and is due to the increase of velocity dispersion at large radius when including the ambiguous stars (middle panel of Figure \ref{fig:fig5}).

\subsection{A deeper look at Segue~I}
Our different procedures return significantly different estimates of $J$ for Seg~I, indicating a lack of robustness against contamination of the stellar-kinematic sample. Let us examine this sensitivity in greater detail.

\subsubsection{Bayesian algorithm for membership probabilities}
\begin{figure}
\begin{center}
\includegraphics[width=\columnwidth]{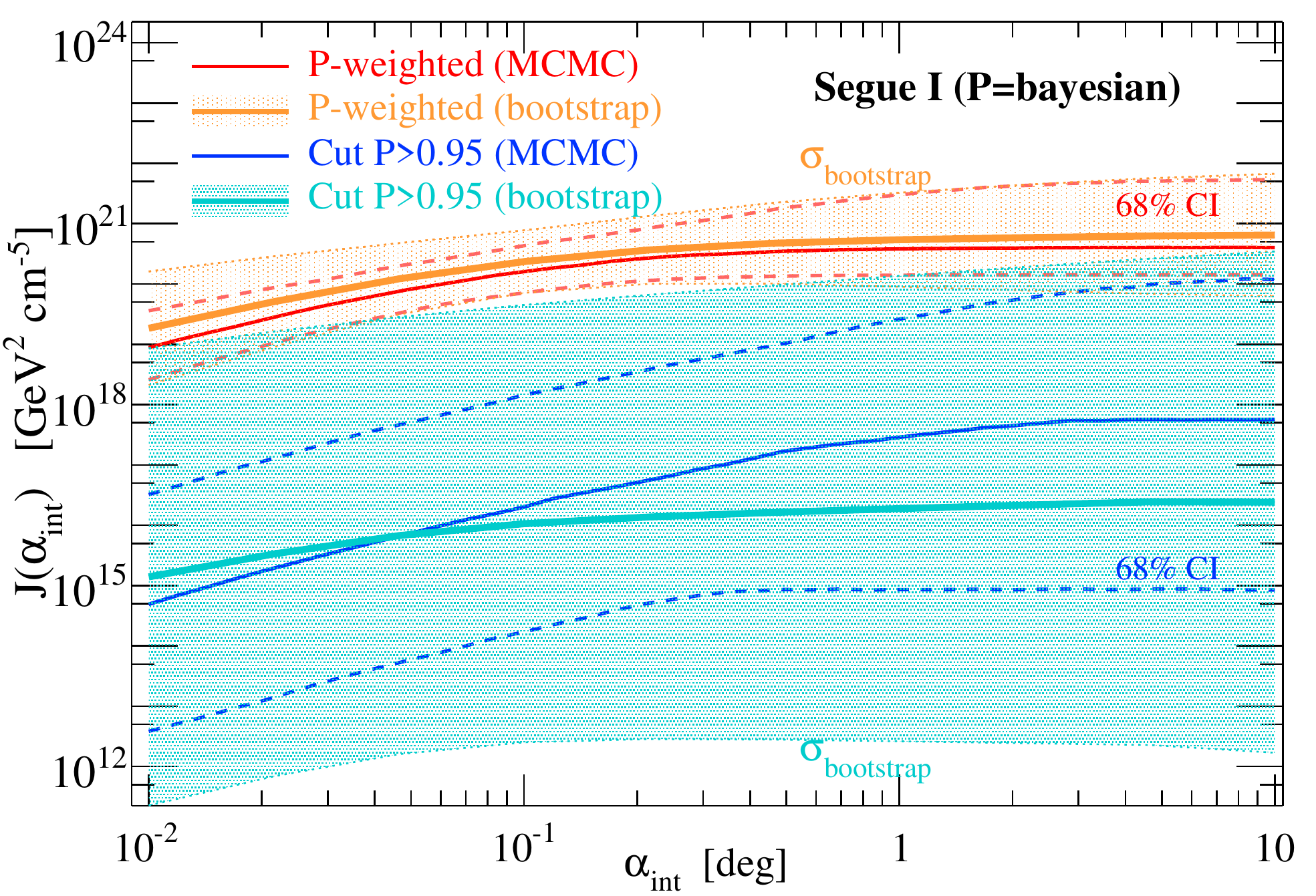}
\includegraphics[width=\columnwidth]{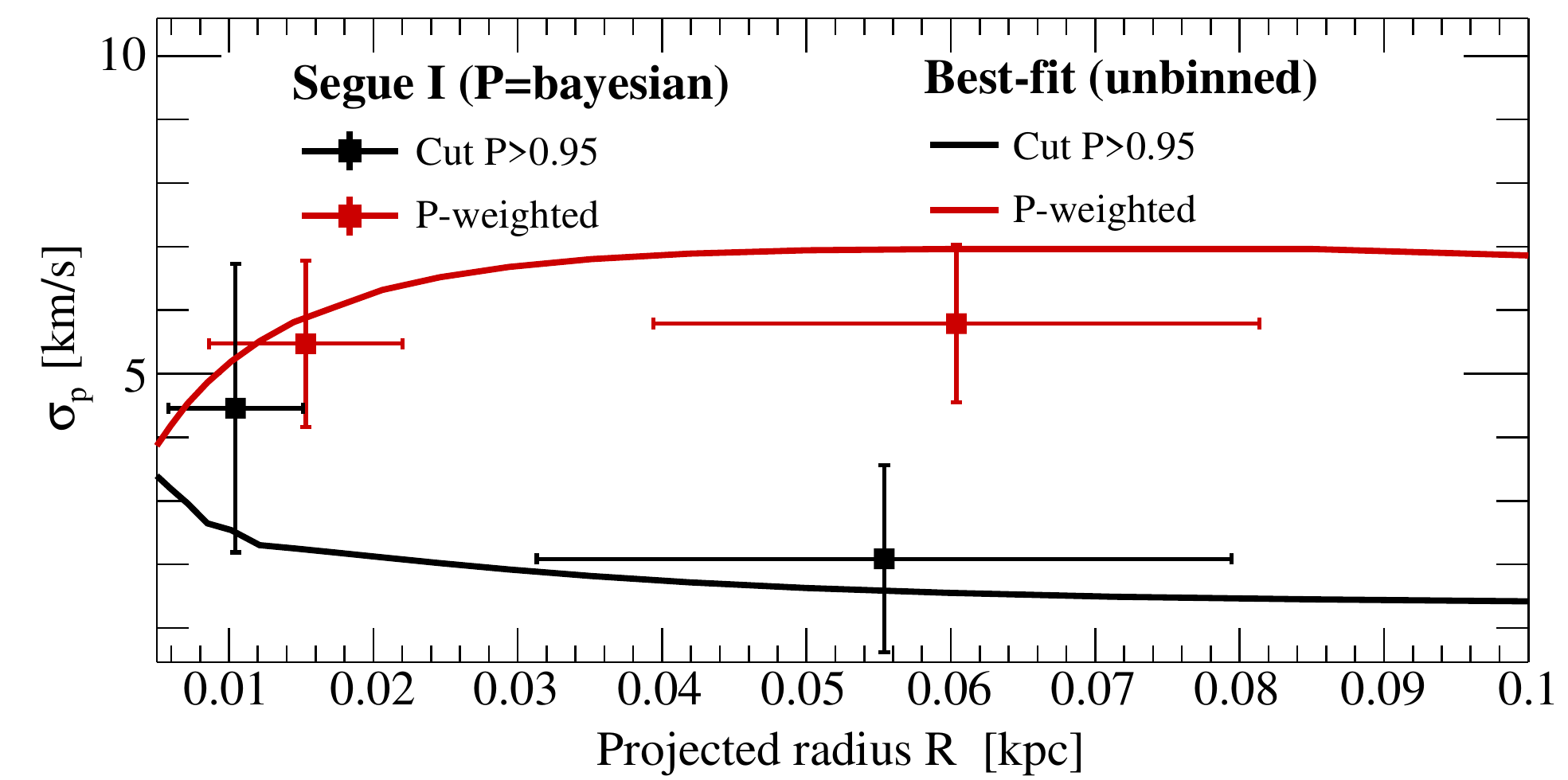}
\includegraphics[width=\columnwidth]{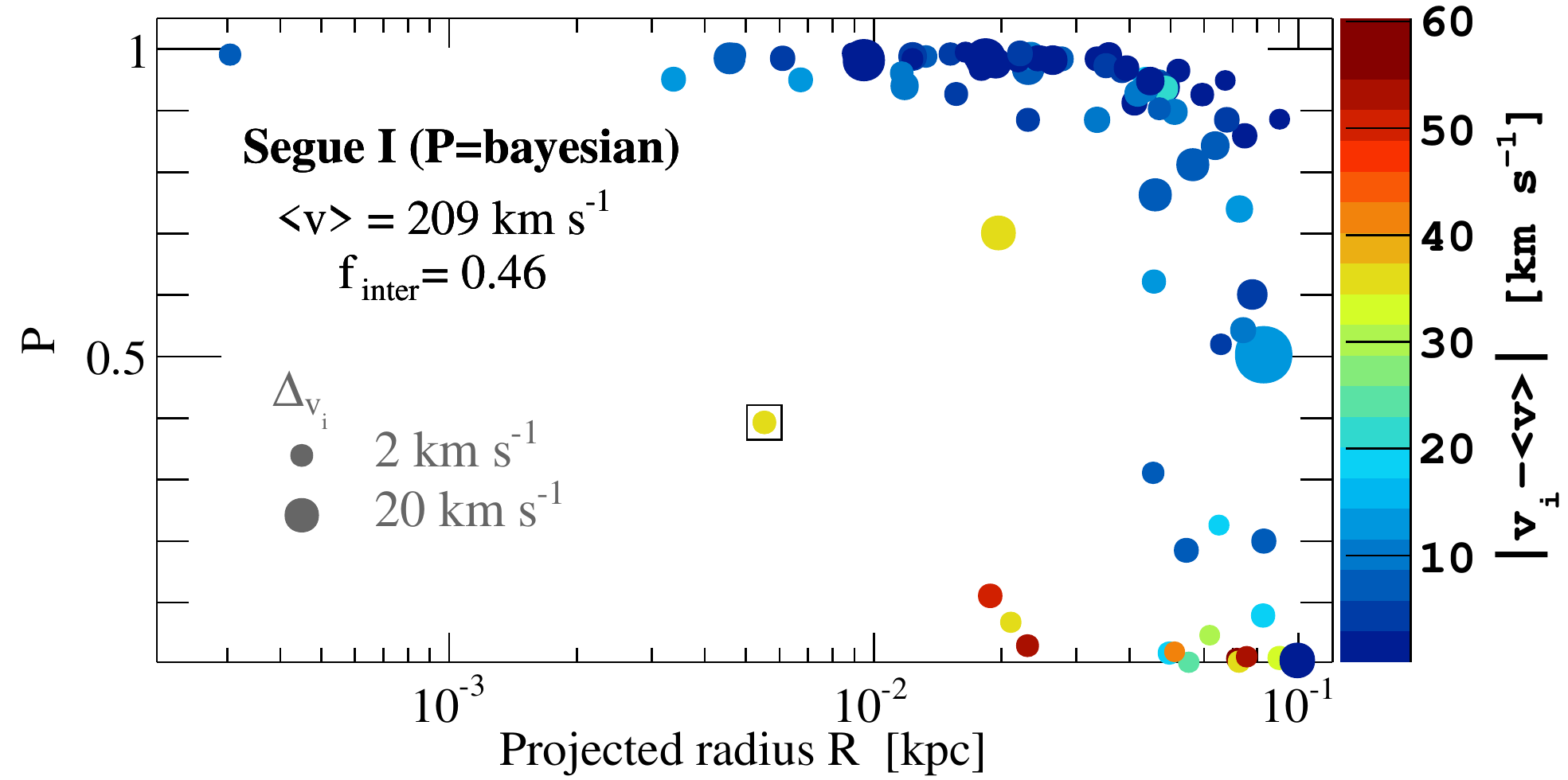}
\caption{Same as in Figure \ref{fig:fig5}, but with the Bayesian estimation of the membership probabilities \citepads{2011ApJ...733...46S,2011ApJ...738...55M}. The black square in the bottom panel highlights a star whose exclusion changes drastically the $J$-factor reconstruction. See text for discussion.}
\label{fig:fig6}
\end{center}
\vspace{-0.3cm}
\end{figure}

First, we repeat our analysis of S11's data for Seg~I, but using the membership probabilities that S11 obtain via the Bayesian analysis described by \citetads{2011ApJ...738...55M}.  This gives a \textit{larger} number of stars with ambiguous membership probability ($f_{\rm inter}$ increases from  $0.19$ to $0.46$; bottom panel of Figure \ref{fig:fig6}). The velocity dispersion profiles obtained with the different procedures are very different (middle panel of Figure \ref{fig:fig6}): the velocity dispersion at large radius, obtained with the cut $P \leq 0.95$, is three times smaller than when using the $P$ values as weights. Accordingly, the differences between $J$-factors from our four procedures become even larger than what we obtained previously (top panel of Figure \ref{fig:fig6}): the $P$-weighted analysis gives a value of $J$ that is more than three orders of magnitude larger than the one obtained using the cut-95\% analysis. Interestingly, the cut-95\% bootstrap analysis yields extremely large uncertainties (green area in the top panel of Figure \ref{fig:fig6}), indicating that the estimation of $J$ may be unduly influenced by the presence of a few outliers in the sample.  

\subsubsection{Sensitivity to membership probability threshold}
\begin{figure}
\begin{center}
\includegraphics[width=\columnwidth]{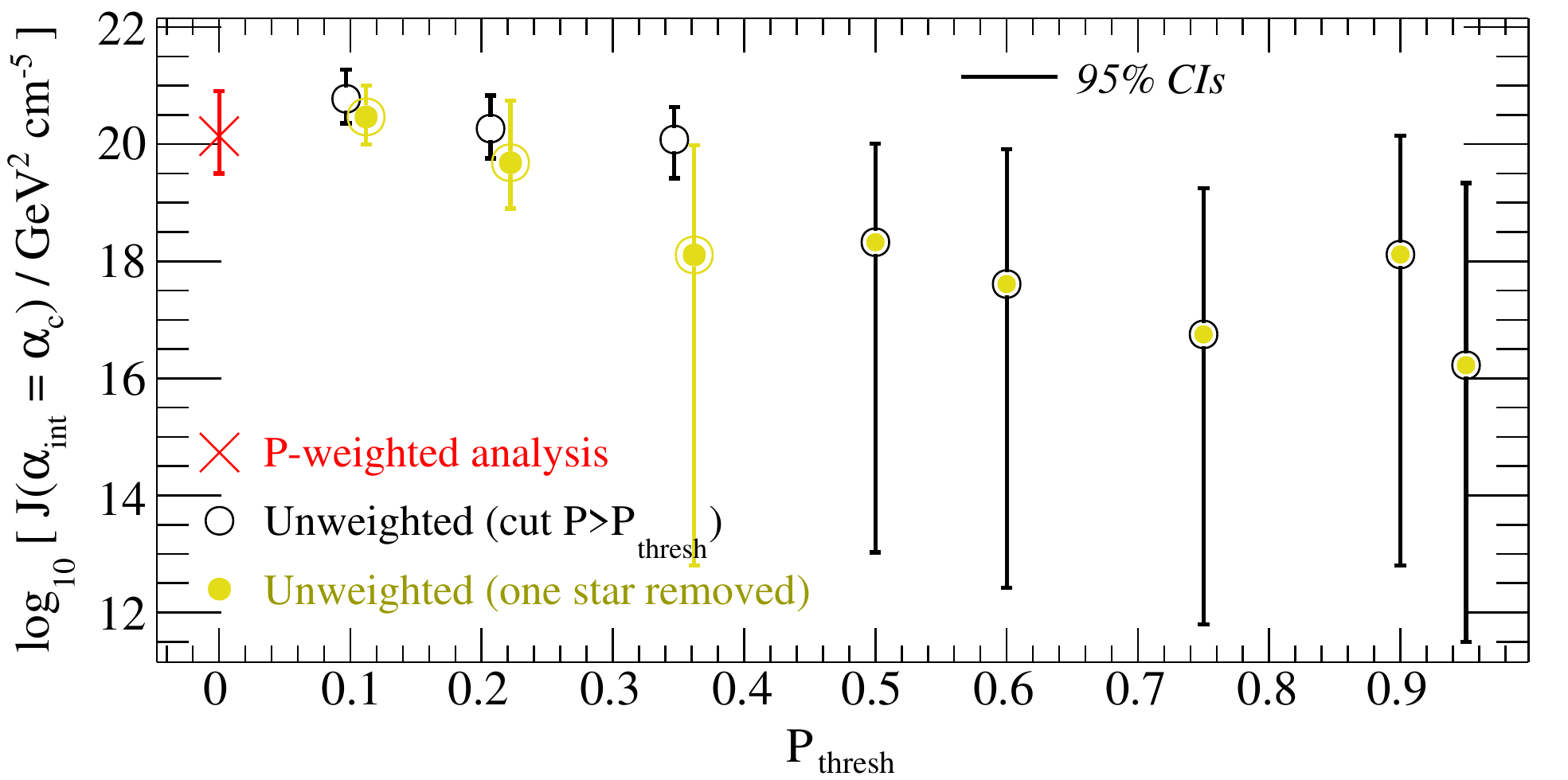}
\caption{$J(\alpha_{\rm int} = \alpha_c = 0.14^\circ$) and 95\% CIs from $P$-weighted (red cross) and $P$-unweighted MCMC analyses as a function of $P_{\rm thresh}$, the threshold for which stars with $P<P_{\rm thresh}$ are discarded from the analysis. The difference between filled yellow circles and black circles is the removal of a single star shown with a similar symbol in the bottom panel of Figure \ref{fig:fig6} (star at $R=5.5$~pc and with $P=0.39$).}
\label{fig:fig7}
\end{center}
\vspace{-0.3cm}
\end{figure}

Our choice of threshold $P_{\rm thresh} = 0.95$ for selecting high-probability members is arbitrary. We now examine the impact of our choice of $P_{\rm thresh}$ on the estimation of Seg~I's $J$-factor, still using our MCMC procedure and S11's Bayesian estimates of membership probabilities. Figure~\ref{fig:fig7} shows how results change as we vary the threshold from $P_{\rm thresh} = 0.1$ to $0.95$.   Thresholds of $P_{\rm thresh}\geq 0.5$ give small $J$-factors with large uncertainties.  Thresolds of $P_{\rm thresh}<0.5$ give larger $J$-factors with much smaller uncertainties, the natural consequence of including more stars with large deviations from the mean velocity.  

S11 themselves note that their estimate of Seg~I's velocity dispersion is sensitive to the inclusion/exclusion of one particular star, SDSSJ100704.35+160459.4, which is a $6\sigma$ velocity outlier but, since it is close to the center of Seg~I, receives $P>0.95$ from the EM algorithm and $P\approx 0.5$ from the Bayesian algorithm.  Following their lead, we excluded this star from all of our analyses (see Section \ref{subsec:seg1}), so it cannot be responsible for the sensitivity we find with regard to choice of procedure.  Instead, we find strong sensitivity to whether we include or exclude another star, SDSSJ100659.95+160408.7\footnote{This star is not among the seven cases of ambiguous membership that S11 discuss in their Appendix.} (yellow dot highlighted by a black square in the bottom panel of Figure~\ref{fig:fig6}), for which the Bayesian algorithm gives $P\approx 0.39$.  For cases with $P_{\rm thresh} < 0.4$, Figure \ref{fig:fig7} illustrates this sensitivity by showing the effect of removing this one star from the sample that is otherwise specified by $P_{\rm thresh}$.  Open black circles show results from analyses that include the star, and yellow circles with rings show results from analyses that exclude it.  The difference can be as large as $\sim 2$ orders of magnitude in $J$. 

\subsubsection{Understanding discrepancies among previously derived $J$-factors}
Several published analyses of Seg~I have used S11's data to estimate $J$-factors that are larger and less uncertain than we have recently reported \citepads[][`B15' hereafter]{2015MNRAS.453..849B}.  In that work, we used exclusively the `cut-95\%' procedure with MCMC estimation of parameters.  In order to understand apparent discrepancies with respect to B15's result for Seg~I, $\log J(0.25^{\circ})=16.8_{-2.1}^{+2.0}$ [GeV$^2$ cm$^{-5}$], we have reproduced two published analyses of Seg~I which use similar techniques.

First, \citetads[][`E10' hereafter]{2010PhRvD..82l3503E} apply a similar Jeans analysis to the 66 stars from S11 for which the EM algorithm gives $P>0.8$, all of which receiving an equal weight.  E10 assume an Einasto profile for the DM density, an isotropic velocity dispersion ($\beta_{\rm ani}= 0$) and a Plummer light profile $((\alpha,\beta,\gamma)=(2,5,0)$ in Eq. \ref{eq:nu}). Compared to our setup, their prior ranges on the DM density are narrower, with e.g. $\alpha \in [0.14:0.3]$ while we use $\alpha \in [0.12:1]$ (see \citealtads{2015MNRAS.453..849B}). E10 report $\log J(0.25^{\circ})=19.1 \pm 0.6$, with CIs based on a fit of a Gaussian function to the posterior PDF that they obtain for $J$.  Relaxing our membership threshold to $P_{\rm thresh}=0.8$ and adopting the same assumptions as E10, we can reproduce their estimate so long as we also follow their procedure.  However, we find that the PDF for $J$ is significantly non-Gaussian, with long tails that extend toward small values.  While a Gaussian fit gives $\log J(0.25^{\circ})=18.8 \pm 0.5$, consistent with E10's result, the median value (with CIs enclosing the central 68\% of probability) of the actual PDF is $\log J(0.25^{\circ})=18.5_{-1.3}^{+0.8}$, leading to much larger uncertainties.  Thus the discrepancy between B15's and E10's estimates of Seg~I's $J$-factor is due to 1) their use of a less restrictive member sample, 2) their fit of a Gaussian function to a non-Gaussian PDF, and 3) their more restrained ranges on DM density priors. 

We have also repeated the analysis of \citetads[][`G-S15' hereafter]{2015ApJ...801...74G}, who use the S11 sample and enforce a membership threshold $P_{\rm thresh}^{\rm Bayes}>0.5$ (G-S15's paper erroneously reports using a threshold of 0.95 for Seg~I; this stricter threshold was used only for their analysis of `classical' dSphs).  We can reproduce G-S15's estimate of $J$ if we adopt their modelling assumptions and use $P_{\rm thresh}=0.5$. We find that the discrepancy between G-S15's estimate of $\log J(0.5^{\circ})=19.36_{-0.35}^{+0.32}$ and B15's is due to 1) different thresholds for membership probabilities, and 2) their choice of anisotropy prior, which favours $\beta_{\rm ani} = 1$ (radial anisotropy), while we employ a flat prior\footnote{\citetads{2015MNRAS.446.3002B} found that the exact parametrisation of $\beta_{\rm ani}$ has no strong impact on the reconstruction of the $J$-factor of `ultrafaint' dSphs, but the shape of the prior can be important.} on $\beta_{\rm ani}$ on the range $[-9:1]$.

\section{Summary}
\label{sec:conc}
In order to help guide efforts at indirect detection of DM particles in dSph galaxies, we have examined sensitivity of expected gamma-ray fluxes to contamination of the stellar-kinematic samples on which estimates of DM densities are based.  Using a large suite of mock stellar-kinematic data sets that include varying degrees of contamination by Galactic foreground and stream populations, we compare estimated to true values of the `$J$-factor', which is proportional to the expected gamma-ray fluxes from DM annihilation.  We find that, for the small stellar-kinematic samples that are available for `ultrafaint' dSph galaxies, contamination by foreground stars can cause $J$-factors to be overestimated by several orders of magnitude.  

In order to gauge the level at which contamination affects estimates of $J$ for real dSphs, we have identified characteristics of mock data sets for which estimates of the $J$-factor are unreliable.  Such data sets tend 1) to have a large fraction of stars with ambiguous membership status, as quantified by estimates of membership probabilities, and 2) to give different results depending on details of arbitrary choices about whether to include or exclude specific stars with ambiguous membership status.  

It is difficult to judge what fraction of the known dSph galaxies are vulnerable to these sorts of complications. For the 8 known `classical' dwarfs we studied in \citetads{2015MNRAS.453..849B}, we find that only Fornax shows specific signs of contamination. We have considered the stellar-kinematic data sets that are available for two of the `ultrafaint' dSph galaxies that have been the focus of several indirect detection investigations: Ret~II and Seg~I. The available samples for Ret~II suggest a relatively small degree of contamination and contain relatively few stars with ambiguous membership status; as a consequence, our estimates of $J$ for Ret~II do not exhibit sensitivity to various possible procedures for addressing membership.  In contrast, the available sample for Seg~I has a larger fraction of foreground contaminants, and a substantially larger fraction of stars with ambiguous membership status. Furthermore, we find that estimates of $J$ for this object can vary by several orders of magnitude depending on the inclusion/exclusion of a few stars with ambiguous membership status. In general, future analyses of dSph galaxies should check explicitly for such sensitivity to interlopers.

\begin{figure}
\begin{center}
\includegraphics[width=\columnwidth]{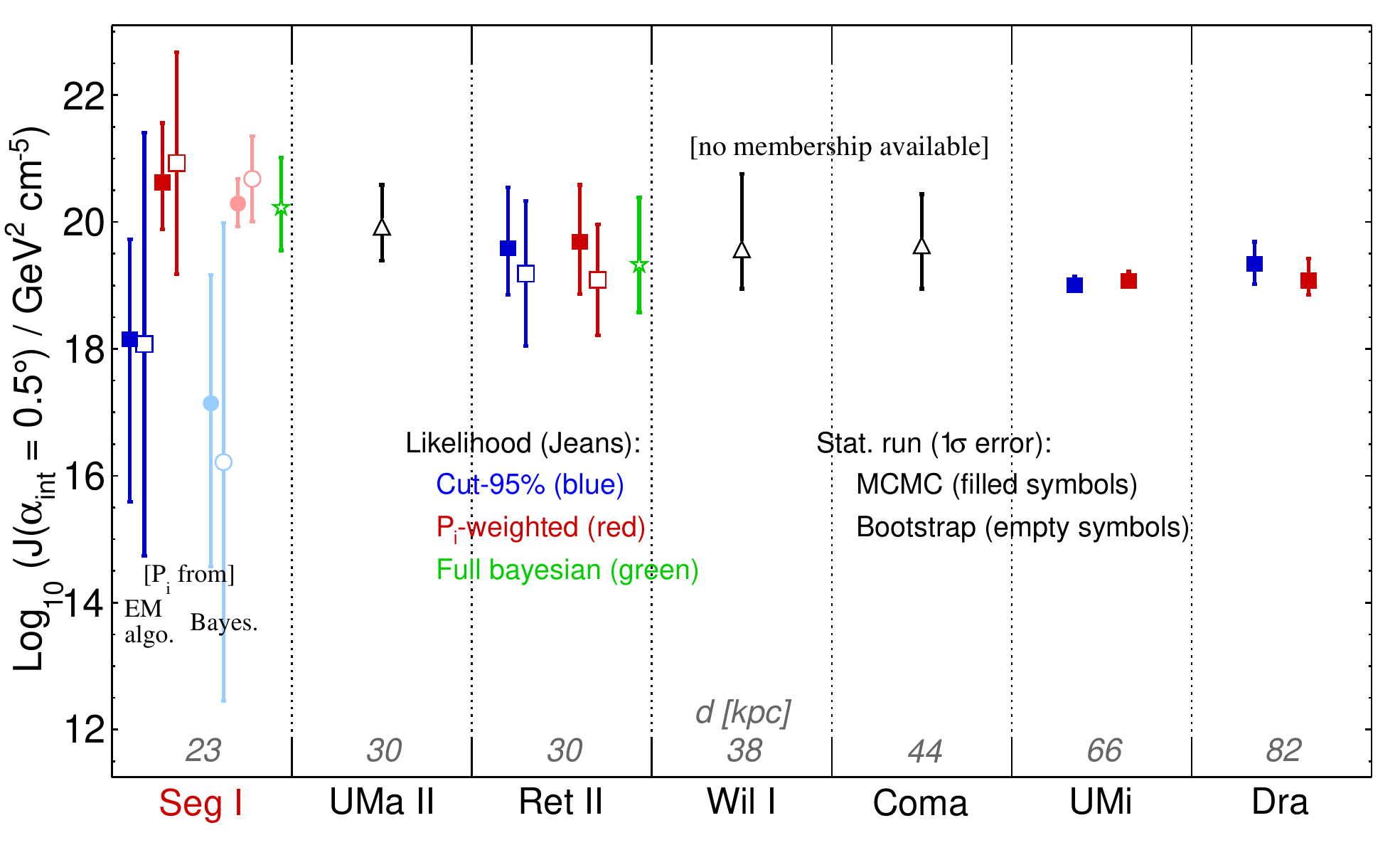}
\caption{Comparison of the $J$-factors of Seg~I, obtained with the several tests of this paper, to the values for the closest dSphs galaxies from \citetads{2015MNRAS.453..849B,2015ApJ...808L..36B}. The $J$-factor can vary from $\sim 10^{16}$ to $\sim 10^{21}$ GeV$^{2}$ cm$^{-5}$ from one analysis to another. On the other hand, Ret~II's $J$-factor was found to be robust against the different tests.}
\label{fig:fig8}
\end{center}
\vspace{-0.3cm}
\end{figure}
Figure \ref{fig:fig8} summarises the sensitivity to contamination that we find when analysing Seg~I's kinematic sample in various ways, and compares the range of estimated $J$-factors to those we obtain for other nearby dSphs \citepads{2015MNRAS.453..849B,2015ApJ...808L..36B}.  For procedures that give non-zero weight to marginal members in the wings of Seg~I's velocity distribution, we obtain $J$-factors that are larger and have relatively small CIs compared to other dSphs; similar results from similar analyses have motivated deep gamma-ray observations of Seg~I (e.g., \citealtads{2012PhRvD..85f2001A,2014JCAP...02..008A}).  On the other hand, procedures that use stricter membership criteria give estimates for $J$ that are significantly smaller and more uncertain than obtained for other dSph galaxies\footnote{For Ursa Major II (UMa II), Willman I (Wil I) and Coma, we could not repeat our tests for sensitivity to choice of procedure for handling membership probabilities, as the relevant stellar-kinematic samples are not publicly available.}.  Depending on choice of procedure, for Seg~I we can recover estimates of $J$-factors spanning $\sim 3$ orders of magnitude, covering the range of previously published values. The results of the more sophisticated, full Bayesian analysis (App.~\ref{subsubsec:mixture}) gives a high $J$ value, but we note that this method appears to suffer from the same (and sometimes worse) drawbacks than the simpler analysis in presence of strong contamination levels. We emphasize however that a more detailed testing of this approach, beyond the scope of this paper, would be necessary to understand the origin of these drawbacks. We conclude that estimates of $J$-factors for Seg~I should be regarded with extreme caution when planning and interpreting indirect detection experiments.  

Further spectroscopic measurements of Seg~I could help resolve this issue, by improving the quality and the amount of data. This task, as well as improving algorithms for separating members from any kind of foregrounds, should receive a high priority given the importance of this object.

\section*{Acknowledgements}
We thank C\'eline Combet and Alex Geringer-Sameth for useful discussions, and Manoj Kaplinghat for comments that have helped us improve the paper. This work has been supported by the ``Investissements d'avenir, Labex ENIGMASS" and by the French ANR, Project DMAstro-LHC,
ANR-12-BS05-0006. MGW is supported by National Science Foundation grants AST-1313045, AST-1412999.  This study used the CC-IN2P3 computation center of Lyon.
\label{lastpage}

\appendix
\section{Alternative procedure for assessing membership and estimating $J$ simultaneously}
\label{subsubsec:mixture}
Both the 'cut-95' and 'P-weighted' analyses suffer from the facts that 1) stellar membership is evaluated under a different model (i.e., one that assumes the velocity dispersion of members is independent of position) than the one that is used subsequently to infer $\rho(r)$, and 2) uncertainties in membership probabilities are not taken into account.  Therefore we also consider an alternative analysis that allows for simultaneous evaluation of membership and inferences about $\rho(r)$ (and hence $J(\alpha_{\rm int})$).  Specifically, we consider a mixture model under which $P(R,V,Z|\vec{\theta})$ represents the joint probability density of stellar position $R$, line-of-sight velocity $V$ and metallicity $Z$, given a model specified by the vector $\vec{\theta}$ of free parameters.  If a star is drawn randomly from a mixture of $N_p$ populations, where (unobserved) variable $M\in\{1,2,...,N_{\rm p}\}$ indicates that star's population, then 
\begin{eqnarray}
  P(R,V,Z|\vec{\theta})=\displaystyle\sum_{M=1}^{\rm N_p} P(R,V,Z,M|\vec{\theta})\nonumber\\
  =\displaystyle\sum_{M=1}^{\rm N_p}P(M|\vec{\theta})P(R,V,Z|M,\vec{\theta})\nonumber\\
  =\displaystyle\sum_{M=1}^{\rm N_p}P(M|\vec{\theta})P(R|M,\vec{\theta}) P(V,Z|M,R,\vec{\theta}).
  \label{eq:jointprob}
\end{eqnarray}
In the first line, we marginalise over the unknown variable $M$.  In each of the last two lines we apply the chain rule of probability: $P(A,B)=P(A|B)P(B)$.  Here, $P(M|\vec{\theta})$ is the probability that the star is drawn from population $M$, given the model.  $P(R|M,\vec{\theta})$ is the probability density of position, given the model and that the star is drawn from population $M$.  $P(V,Z|M,R,\theta)$ is the joint probability density of velocity and metallicity, given the model and that the star is drawn from population $M$ and has position $R$.  

The probability density $P(R|M,\vec{\theta})$ can be written in terms of projected stellar density profile, $I_M(R)$, for the population specified by $M$.  Assuming spherical symmetry, a circle of radius $R$ encloses $N_M(R)=2\pi\int_0^R\,S\,I_M(S)\,d S$ of this population's stars as seen in projection, and 
\begin{eqnarray}
  P(R|M,\vec{\theta})\!\!\!&\!=\!&\!\!\!\frac{d}{d R}\biggl [ \frac{N_M(R)}{N_{M}(\infty)}\biggr ]
  =\frac{d}{d R}\biggl [\frac{2\pi\int_{0}^{R}S\,I_M(S)\,d S  }{N_M(\infty)}\biggr ]\nonumber\\
 \!\!\! &\!=\!&\!\!\!\frac{2\pi R\,I_M(R)}{N_M(\infty)}.
  \label{eq:pr}
\end{eqnarray}
The probability $P(M|\vec{\theta})$ is the fraction, $f_M$, of stars that belong to population $M$:
\begin{equation}
  P(M|\vec{\theta})=\frac{N_M(\infty)}{N_{\rm tot}}\equiv f_M,
  \label{eq:pm}
\end{equation}
where $N_{\rm tot}\equiv N_1(\infty)+N_2(\infty)+...+N_{\rm p}(\infty)$ is the total number of stars, including all populations.  Substituting Eqs.~(\ref{eq:pr}) and~(\ref{eq:pm}) into Eq.~(\ref{eq:jointprob}), we obtain
\begin{equation}
  P(R,V,Z|\vec{\theta})=\frac{2\pi R}{N_{\rm tot}}\displaystyle\sum_{M=1}^{N_{\rm p}}I_{M}(R)P(V,Z|R,M,\vec{\theta}).
\end{equation}
    
Then a data set consisting of independent observations of position, velocity and metallicity for $N$ stars has likelihood
\begin{equation}
  \mathcal{L}=\displaystyle\prod_{j=1}^{N}\biggl (\frac{2\pi R_j}{N_{\rm tot}}\displaystyle\sum_{M=1}^{N_{\rm p}}I_{M}(R_j)P(V_j,Z_j|R_j,M,\vec{\theta})\biggr ).
  \label{eq:likelihood}
\end{equation}
In order to analyse stellar-kinematic data sets for dwarf spheroidals in the presence of foreground contamination, here we assume $N_{\rm p}=2$, allowing for member ($M=1$) and foreground ($M=2$) populations.  We model the 3D stellar density using Eq.~(\ref{eq:nu}). We assume the stellar density of foreground stars is uniform over the relatively small field subtended by the member population, such that the foreground population has projected density $I_2$, a constant.  Then the fraction of observed stars that are members is 
\begin{equation}
  f_{1}=\frac{4\pi}{N} \displaystyle\int_0^{\infty}r^2\nu_1(r)d r=\frac{4\pi k_1 r_1^3}{\alpha_1 N}\frac{\Gamma\bigl (\frac{\beta_1-3}{\alpha_1}\bigr )\Gamma\bigl (\frac{3-\gamma_1}{\alpha_1}\bigr )}{\Gamma\bigl (\frac{\beta_1-\gamma_1}{\alpha_1} \bigr )}
  \label{eq:f1}
\end{equation}
where $\Gamma(x)$ is the gamma function.  The fraction of observed stars that are foreground is
\begin{equation}
  f_{2}=1-f_1=\frac{\pi R^2_{\rm max} I_2}{N},
  \label{eq:f2}
\end{equation}
where $R_{\rm max}$ is the maximum radius of the (circular) field.  

Finally, we assume that the velocities and metallicities of both member and foreground populations follow bivariate normal distributions,
\begin{equation}
  P(V,Z|M,R,\vec{\theta})=\mathcal{N}_2(\vec{\mu}_M,\vec{\Sigma}_M),\\
\end{equation}
with mean vectors
\begin{equation}
  \vec{\mu}_M\equiv
  \left(
    \begin{matrix}
      \langle{V}\rangle_M\\
      \langle{Z}\rangle_M
    \end{matrix}
  \right )
  \label{eq:mu1}
\end{equation}
and covariance matrices
\begin{equation}
  \vec{\Sigma}_M\equiv
\left(
\begin{matrix}
  \sigma^2_{V,M}(R)+\delta^2_{V} & 0\\
  0 & \sigma^2_{Z,M}+\delta^2_{Z}
\end{matrix}
\right )
\label{eq:covmatrix}
\end{equation}
whose diagonal elements are broadened by observational errors $\delta_{V}$ and $\delta_Z$ (we implicitly assume that velocity and metallicity are uncorrelated).  We assume the velocity dispersion of the foreground population is uniform across the field of view, but we allow the velocity dispersion of the member population to vary with position.  Specifically, we use Eq.~(\ref{eq:einasto}) to model $\rho(r)$, and then use the Jeans equation to compute the variance of the projected velocity distribution as described in Section \ref{subsec:MCMC}.  

Our mixture model is fully specified by 18 free parameters: 
\begin{equation}
  \vec{\theta}=
  \left(
    \begin{matrix}
      \log_{10}[\rho_s/(M_{\odot}/\mathrm{pc}^3)]\\
      \log_{10}[r_s/\mathrm{pc}]\\
      \alpha\\
      \beta\\
      \gamma\\
      -\log_{10}[1-\beta_a]\\
      f_1\\
      \log_{10}[r_1/\mathrm{pc}]\\
      \alpha_1\\
      \beta_1\\
      \gamma_1\\
      \langle V\rangle_1\\
      \langle V\rangle_2\\
      \langle Z\rangle_1\\
      \langle Z\rangle_2\\
      \sigma^2_{V,2}\\
      \sigma^2_{Z,1}\\
      \sigma^2_{Z,2}
    \end{matrix}
  \right ).
  \label{eq:params}
\end{equation}
\bibliography{bmw}
\end{document}